\documentstyle[eqsecnum,pre,aps,epsf,epsfig]{revtex}

\begin{document}
\draft

\newcommand{\argu}{\scriptstyle {\rm{\bf p}}, \stackrel{\mbox{{\tiny $ \leftrightarrow 
$}}}{\lambda} \textstyle}
\newcommand{\argum}{\scriptstyle -{\rm{\bf p}}, -\stackrel{\mbox{{\tiny $ 
\leftrightarrow $}}}{\lambda} \textstyle}
\newcommand{\k}{\big\{ \stackrel{\mbox{{\tiny $ \leftrightarrow $}}}{\kappa} \big\}}
\newcommand{\argi}{\scriptstyle {\rm{\bf p}}_i, \stackrel{\mbox{{\tiny $ \leftrightarrow 
$}}}{\lambda}_i \textstyle}
\newcommand{\argun}{\scriptstyle {\rm{\bf p}} \textstyle}
\newcommand{\argumn}{\scriptstyle -{\rm{\bf p}} \textstyle}
\newcommand{\n}{\big\{ \stackrel{\mbox{{\tiny $ \leftrightarrow $}}}{0} \big\}}
\newcommand{\argin}{\scriptstyle {\rm{\bf p}}_i \textstyle}
\newcommand{\arga}{\scriptstyle {\rm{\bf p}}_1, \stackrel{\mbox{{\tiny $ \leftrightarrow 
$}}}{\lambda}_1 \textstyle}
\newcommand{\argb}{\scriptstyle {\rm{\bf p}}_n, \stackrel{\mbox{{\tiny $ \leftrightarrow 
$}}}{\lambda}_n \textstyle}

\title{Noisy random resistor networks: renormalized field theory for the multifractal moments of the current distribution
}
\author{Olaf Stenull and Hans-Karl Janssen
}
\address{
Institut f\"{u}r Theoretische Physik 
III\\Heinrich-Heine-Universit\"{a}t\\Universit\"{a}tsstra{\ss}e 1\\40225 
D\"{u}sseldorf, Germany
}
\date{\today}
\maketitle

\begin{abstract}
We study the multifractal moments of the current distribution in randomly diluted resistor networks near the percolation treshold. When an external current is applied between to terminals $x$ and $x^\prime$ of the network, the $l$th multifractal moment scales as $M_I^{(l)} \left( x, x^\prime \right) \sim \left| x - x^\prime \right|^{\psi_l /\nu}$, where $\nu$ is the correlation length exponent of the isotropic percolation universality class. By applying our concept of master operators [Europhys.\ Lett.\ {\bf 51}, 539 (2000)] we calculate the family of multifractal exponents $\left\{ \psi_l \right\}$ for $l \geq 0$ to two-loop order. We find that our result is in good agreement with numerical data for three dimensions.
\end{abstract}

\section{Introduction}
\label{noisyIntroduction}
Percolation\cite{percolationReviews} is a leading paradigm for disorder. It provides an intuitively appealing and transparent model of the irregular geometry which occurs in disordered systems. Moreover, it is a prototype of a phase transition. Though percolation represents the simplest model of a disordered system, it has many applications, e.g., polymerization, porous and amorphous materials, thin films, spreading of epidemics etc.  

In particular the transport properties of percolation clusters have gained a vast amount of interest over the last decades. Random resistor networks (RRN) are a prominent model for transport on percolation clusters. By means of RRN one can study the conductivity of disordered media which might be important for technical applications. Nonlinear random resistor networks, for which the voltage drop over an individual resistor depends on some power of the current flowing through it, can be exploited to derive various fractal dimensions of percolation clusters. From the conceptual point of view, RRN have the advantage that one can formulate a field theoretic Hamiltonian amenable to renormalization group analysis. Via RRN one can learn about diffusion on disordered substrates, since the diffusion constant $D$ and the conductivity $\Sigma$ of the system are related by the Einstein relation 
\begin{eqnarray}
\label{einRel}
\Sigma = \frac{e^2 n}{K_B T} D \ ,
\end{eqnarray}
where $e$ and $n$ denote the charge and the density of the mobile particles. The connection of the two problems is particularly important, since up to date no direct approach to diffusion on percolation clusters by means of a dynamic field theory exists.

In this paper we study the distribution of currents in RRN. The current distribution has many interesting features, one of which is multifractality\cite{evertsz_mandelbrot_92}. This means that the distribution is not controlled by one or two relevant length scales, but rather by an infinite hierarchy of such length scales. The concept of multifractality was introduced for turbulence\cite{mandelbrot_74}. It has been applied successfully in diverse areas including diffusion near fractals\cite{cates_witten_87}, electrons in disordered media\cite{wegner_87}, polymers in disordered media\cite{gersappe_etal_91}, random ferromagnets\cite{ludwig_87}, chaotic dissipative systems\cite{halsey_etal_86}, and heartbeat\cite{ivanov_99}.

Due to the multifractality infinitely many exponents are needed to characterize the current distribution. Consider two connected terminals $x$ and $x^\prime$ of the network. Suppose a current $I$ is inserted at $x$ and withdrawn at $x^\prime$. The $l$th moment of the current distribution given (apart from technical details, c.f., Sec.~\ref{momentsOfTheCurrentDistribution}) by,  
\begin{eqnarray}
M_I^{(l)} \left( x, x^\prime \right) = \left\langle \sum_b  i_b^{2l} \right\rangle_C \ ,
\end{eqnarray}
where the sum runs over all current carrying bonds (the backbone), $\langle \cdots \rangle_C$ stands for the average over all diluted configurations and $i_b$ is an abbreviation for $I_b /I$, scales at criticality as~\cite{rammal_etal_85}
\begin{eqnarray}
\label{scalingMultifracMom} 
M_I^{(l)} \left( x, x^\prime \right) \sim \left| x - x^\prime \right|^{-x_l} \
.
\end{eqnarray}
The $x_l$ constitute an infinite set of exponents which are not related to each other in a linear fashion, i.e., the multifractal moments do not show the usual gap scaling commonly encountered in critical phenomena.

Each of the $M_I^{(l)}$ is associated with a particular subset of backbone bonds having its distinct fractal dimension. Let $n(i)$ be the number of bonds carrying current $i$. Upon applying the saddle point method on finds that the main contribution to the $l$th moment is given by\cite{deArcangelis&co_86}
\begin{eqnarray}
n \left( i_l \right) \sim \left| x - x^\prime \right|^{f(l)} \
,
\end{eqnarray}
with the multifractal spectrum $f(l)$ and the multifractal exponents $x_l$ being related to each other by a Legendre transformation. $f(l)$ can be interpreted as the fractal dimension of the subset of bonds dominating $M_I^{(l)}$.

An elegant approach for studying the multifractal moments is to consider RRN with microscopic noise, i.e., random networks in which the conductances of the individual resistors fluctuate about some mean. These noisy RRN were originally introduced by Rammal {\em et al}.\cite{rammal_etal_85} to study the effects of Flicker ($1/f$) noise. Flicker noise refers to the low frequency spectrum of excess voltage fluctuations measured when a constant current is applied to a resistor. The $l$th noise cummulant $C^{(l)}_R \left( x, x^\prime \right)$ of the total resistance between the terminals $x$ and $x^\prime$ is proportional to $M_I^{(l)} \left( x, x^\prime \right)$ by virtue of Cohn's theorem\cite{cohn_50}.

Historically, the existence of the set of multifractal exponents $\left\{ x_l \right\}$ was proposed by Rammal {\em et al}\cite{rammal_etal_85}. The authors determined several of their exponents for two dimension by numerical simulations. A set of exponents $\left\{ \zeta_{2l} \right\}$ equivalent to $\left\{ - x_l \nu \right\}$, where $\nu$ is the correlation length exponent for percolation, was also proposed by Arcangelis {\em et al.}\cite{arcangelis_etal_85}. Theses authors derived their exponents for several hierarchical structures analytically. The field theoretic description of multifractality in RRN was pioneered by Park, Harris and Lubensky (PHL)\cite{park_harris_lubensky_87}. Based on an approach by Stephen\cite{stephen_78} they formulated a $D \hspace{-1mm}\times \hspace{-1mm}E$-fold replicated Hamiltonian for noisy RRN. The contributions to the Hamiltonian leading to multifractal behavior contain powers of replica space gradients analog to powers of real space gradients, which were accounted for as an origin of multifractality by Duplantier and Ludwig\cite{duplantier_ludwig_91}. PHL introduced a set of exponents $\left\{ \psi_n \right\}$ identical to the set $\left\{ - x_n \nu \right\}$ and calculated it to first order in $\epsilon = 6-d$, where $d$ denotes the spatial dimension. Later on Fourcade and Tremblay\cite{fourcade_tremblay_95} gave a reinterpretation of the work by PHL. Batrouni {\em et al.}\cite{batrouni&co_96} computed several multifractal exponents for $d=3$ by numerically solving Kirchhoff's equations. Recently Barth\'{e}l\'{e}my {\em et al.}\cite{barthelemy&co_2000} performed simulations indicating that in the thermodynamic limit the $M_I^{(l)}$ do not exist for $l<0$.

In this article we study the moments of the current distribution by renormalized field theory. We extend our real-world interpretation of Feynman diagrams\cite{stenull_janssen_oerding_99,janssen_stenull_oerding_99,janssen_stenull_99} to RRN with noise. Upon introducing multifractal moments for Feynman diagrams we reformulate the field theory of PHL in a way that to our opinion is less complex and more intuitive. By carefully analyzing the relevance of the field theoretic operators related to the noise cumulants, we show that the multifractality is associated with dangerously irrelevant master operators\cite{stenull_janssen_epl2000}. We calculate the set $\left\{ \psi_l \right\}$ for $l\geq 0$ to second order in $\epsilon$. Finally, we compare our result to numerical simulations.

\section{The Model}
\label{noisyModel}
This section provides background on noisy RRN. It is guided by the work of 
Stephen\cite{stephen_78} and PHL\cite{park_harris_lubensky_87}. 

\subsection{Random resistor networks}
\label{noisyRRN}
Consider a $d$-dimensional lattice, where bonds between nearest neighboring sites $i$ 
and $j$ are randomly occupied with probability $p$ or empty with probability $1-p$. Each 
occupied bond $\langle i,j \rangle$ has a conductance $\sigma_{i,j}$. Unoccupied bonds 
have conductance zero. The bonds obey Ohm's law
\begin{eqnarray}
\sigma_{i,j} \left( V_j - V_i \right) = I_{i,j} \ , 
\end{eqnarray}
where $I_{i,j}$ is the current flowing through the bond from $j$ to $i$ and $V_i$ is the 
potential at site $i$.

Suppose a current $I$ is injected into a cluster at site $x$ and withdrawn at site 
$x^\prime$. The union of all sites belonging to all self avoiding paths between $x$ and 
$x^\prime$ is refered to as the backbone between $x$ and $x^\prime$. The power dissipated 
on 
the backbone is by definition
\begin{eqnarray}
\label{powerDef}
P=I \left( V_x - V_{x^\prime} \right) \ .
\end{eqnarray}
Using Ohm's law, it may be expressed entirely in terms of voltages as
\begin{eqnarray}
\label{noisyPowerInTermsOfV}
P=  R (x ,x^\prime)^{-1} \left( V_x - V_{x^\prime} \right)^2 = \sum_{\langle i,j 
\rangle} 
\sigma_{i,j} \left( V_i - V_j \right)^2 =  P \left( \left\{ V \right\} \right) \ .
\end{eqnarray}
Here $R (x ,x^\prime)$ is the total resistance of the backbone, the sum is taken over 
all 
nearest neighbor pairs on the cluster and $\left\{ V \right\}$ denotes the corresponding 
set of voltages. As a consequence of the variation principle
\begin{eqnarray}
\label{noisyVariationPrinciple1}
\frac{\partial}{\partial V_i} \left[ \frac{1}{2} P \left( \left\{ V \right\} 
\right) - \sum_j I_j V_j \right] = 0 \ ,
\end{eqnarray}
one obtains Kirchhoff's law
\begin{eqnarray}
\label{noisyCirquitEquations}
\sum_{\langle j \rangle} \sigma_{i,j} \left( V_i - V_j \right) = - \sum_{\langle j 
\rangle} I_{i,j} =I_i \ ,
\end{eqnarray}
where $I_i = I \left( \delta_{i,x} - \delta_{i,x^\prime} \right)$ and the summations 
extend over the nearest neighbors of $i$.

Alternatively to Eq.~(\ref{noisyPowerInTermsOfV}) the power can by rewritten in terms of 
the currents as
\begin{eqnarray}
\label{noisyPowerInTermsOfI}
P=  R (x ,x^\prime) I^2 = \sum_b \rho_b I_b^2 =  P 
\left( \left\{ I_b \right\} \right) \ ,
\end{eqnarray}
with $\left\{ I_b \right\}$ denoting the set of currents flowing through the 
individual bonds, $b= \langle i,j \rangle$,  and $\rho_b = \sigma_b^{-1}$. Obviously the 
cluster may contain closed loops as subnetworks. 
Suppose there are currents $\left\{ I^{(\mbox{\scriptsize{loop}})} \right\}$ circulating independently 
around a complete set of independent closed loops. Then the power is not only a function 
of $I$ but also of the set of loop currents. The potential drop around closed loops is 
zero. This gives rise to the variation principle
\begin{eqnarray}
\label{noisyVariationPrinciple2}
\frac{\partial}{\partial I^{(\mbox{\scriptsize{loop}})}} P \left( \left\{ I^{(\mbox{\scriptsize{loop}})} \right\} , I \right) 
= 0 \ .
\end{eqnarray} 
Eq.~(\ref{noisyVariationPrinciple2}) may be used to eliminate the loop currents and 
thus provides us with a method to determine the total resistance of the backbone 
via Eq.~(\ref{noisyPowerInTermsOfI}).

Since the resistance of the backbone depends on the configurations $C$ of the randomly 
occupied bonds, one introduces an average $\langle \cdots \rangle_C$ over these 
configurations. It is important to recognize that the resistance between disconnected 
sites is infinite. Therefore one considers only those sites $x$ and $x^\prime$ known to 
be on the same cluster. Practically this is done by introducing the indicator function 
$\chi (x ,x^\prime)$ which, for a given configuration $C$, is unity if $x$ and 
$x^\prime$ 
are connected and zero otherwise. Then the $l$th moment of the resistance $R$ with 
respect to the average $\langle \cdots \rangle_C$ subject to $x$ and $x^\prime$ being on 
the same cluster is given by
\begin{eqnarray}
\left\langle \chi (x ,x^\prime) R (x ,x^\prime )^l \right\rangle_C / \left\langle \chi 
(x 
,x^\prime) \right\rangle_C \ . 
\end{eqnarray}

\subsection{Noise in random resistor networks}
\label{noiseInNoisyRRN}
In the following we consider RRN with noise in the sense that the conductances 
$\sigma_b$ 
of occupied bonds fluctuate about some mean. To be specific, the 
$\sigma_b$ are equally and independently distributed random variables with mean 
$\overline{\sigma}$ and higher cumulants $\Delta^{(l\geq 2)}$. The distribution function 
$f$ might for example be Gaussian. Nevertheless, our considerations are not limited to 
this particular choice. In order to suppress unphysical negative conductances, the 
assumption $\Delta^{(l)} \ll \overline{\sigma}^l$ is made. In general the backbone 
resistance will depend on the set of conductances of occupied bonds $\left\{ \sigma_b 
\right\}$. Its noise average will be denoted by
\begin{eqnarray}
\label{defOfNoiseAverage}
\left\{ R (x ,x^\prime ) \right\}_f = \int \prod_b d \sigma_b f \left( \sigma_b \right) 
R 
(x ,x^\prime ) 
\end{eqnarray}
and the corresponding cumulants by
\begin{eqnarray}
\label{defOfNoiseCumulant}
\left\{ R (x ,x^\prime )^l \right\}^{(c)}_f = \left. \frac{\partial^l}{\partial 
\lambda^l} \ln \left\{ \exp \left[ \lambda R (x ,x^\prime ) \right] \right\}_f 
\right|_{\lambda =0} \ .
\end{eqnarray}

Both kinds of disorder, the random dilution of the lattice and the fluctuation of the 
bond conductances about their mean $\overline{\sigma}$, influence the statistical 
properties of the backbone resistance. They are reflected by the moments
\begin{eqnarray}
\label{defMoment}
M_R^{(l)}(x ,x^\prime) = \left\langle \chi (x ,x^\prime) \left\{ R (x ,x^\prime )^l 
\right\}_f \right\rangle_C / \left\langle \chi (x ,x^\prime) \right\rangle_C 
\end{eqnarray}
and the cumulants
\begin{eqnarray}
\label{defCumulant}
C_R^{(l)}(x ,x^\prime) = \left\langle \chi (x ,x^\prime) \left\{ R (x 
,x^\prime )^l \right\}_f^{(c)} \right\rangle_C / \left\langle \chi (x ,x^\prime) 
\right\rangle_C \ . 
\end{eqnarray}

\subsection{Moments of the current distribution}
\label{momentsOfTheCurrentDistribution}
The noise cumulants $C_R^{(l)}$ characterize the distribution of currents flowing 
through 
the network. This section provides a relation between the $C_R^{(l)}$ and the moments of 
the current distribution.

Equation~(\ref{defOfNoiseAverage}) defines the noise average as an average with respect 
to the distribution of the bond conductances $\sigma_b$. Equally well one might express 
the backbone resistance in terms of the bond resistances and average over the 
distribution of the $\rho_b$. Since the $\sigma_b$ are independently and equally 
distributed, the $\rho_b$ are distributed by the same means. Assume that 
the 
distribution function of the deviations $\delta \rho_b = \rho_b - \overline{\rho}$ of 
the 
resistance of each bond from its average $\overline{\rho}$ has the form
\begin{eqnarray}
g_s \left( \delta \rho_b \right) = \frac{1}{s} h \left( \frac{\delta \rho_b}{s} \right)
\end{eqnarray}
and that
\begin{eqnarray}
\lim_{s\to 0} g_s \left( \delta \rho_b \right) = \delta \left( \delta \rho_b \right) \ .
\end{eqnarray}
$s$ is a variable with units of resistance which sets the scale of the distribution. 
With 
this form of $g_s$, the $n$th cumulant $v_n$ of $\delta \rho_b$ tends to zero as $s^n$. 
This follows from the generating function $c \left( \lambda s \right)$ of the $v_n$:
\begin{eqnarray}
\exp \left[ c \left( \lambda s \right) \right] = \left\{ \exp \left( \lambda \delta 
\rho_b \right) \right\}_f = \int dy \ h \left( y \right) \exp \left( \lambda s y \right) 
= \exp 
\left( \sum_{n=1}^\infty \frac{\lambda^n}{n!} v_n\right) \ ,
\end{eqnarray}
where $v_n = c_n s^n$ with $c_n$ being constants. In general $\left\{ R (x ,x^\prime )^l 
\right\}_f^{(c)}$ depends on the entire set of cumulants $\left\{ v_n \right\}$. 
However, 
in the limit $s\to 0$ the leading term is proportional to $v_l$  as we will see 
immediately. Consider the generating function $C \left( \lambda \right)$ of the 
cumulants 
$\left\{ R (x ,x^\prime )^l \right\}_f^{(c)}$, 
\begin{eqnarray}
\exp \left[ C \left( \lambda \right) \right] = \int \prod_b d \delta \rho_b \ g_s \left( 
\delta \rho_b \right) \exp \left[ \lambda R (x ,x^\prime ) \right] \ .
\end{eqnarray}
Expansion of the backbone resistance in a power series in the $\delta \rho_b$ leads to
\begin{eqnarray}
\label{schubidu}
\exp \left[ C \left( \lambda \right) \right] = \int \prod_b d y_b \ h \left( y_b \right) 
\exp \left[ \lambda R^0 (x ,x^\prime ) + \lambda \sum_{k=1}^\infty \sum_{b_1 , \cdots , 
b_k} \frac{s^k}{k!} \left. \frac{\partial^k R (x ,x^\prime )}{\partial \rho_{b_1} \cdots 
\partial \rho_{b_k}} \right|_{\overline{\rho}} y_{b_1} \ldots y_{b_k} \right] \ ,
\end{eqnarray}
where $R^0 (x ,x^\prime )$ is the resistance when $\delta \rho_b = 0$ for every bond 
$b$. 
Equation~(\ref{schubidu}) can be rearranged as
\begin{eqnarray}
\exp \left[ C \left( \lambda \right) \right] &=& \exp \left[ \lambda R^0 (x ,x^\prime ) 
+ 
\lambda \sum_{k=2}^\infty \sum_{b_1 , \cdots , b_k} \frac{s^k}{k!} \left. 
\frac{\partial^k R (x ,x^\prime )}{\partial \rho_{b_1} \cdots \partial \rho_{b_k}} 
\right|_{\overline{\rho}} \frac{\partial^k }{\partial z_{b_1} \cdots \partial z_{b_k}}
\right] 
\nonumber \\
&\times& \prod_b \exp \left[ c \left( z_b \right) \right] \Bigg|_{\lambda s \sum_b 
\left. 
\frac{\partial R (x ,x^\prime )}{\partial \rho_b} \right|_{\overline{\rho}}}
\nonumber \\
&=& \exp \left[ \lambda R^0 (x ,x^\prime ) + \sum_{l=1}^\infty \left( \lambda s 
\right)^l 
c_l \sum_b \left(  \left. \frac{\partial R (x ,x^\prime )}{\partial \rho_b} 
\right|_{\overline{\rho}} \right)^l + \sum_{i=2}^\infty f_i \left( \lambda s^i \right)  
\right] \ ,
\end{eqnarray}
where $f_i$ are functions of $\lambda s^i$. Hence for $l\geq 2$,
\begin{eqnarray}
\left\{ R (x ,x^\prime )^l \right\}_f^{(c)} = c_l \sum_b \left( s \left. 
\frac{\partial R (x ,x^\prime )}{\partial \rho_b} \right|_{\overline{\rho}} \right)^l  
\Big( 1 + {\em O} ( s ) \Big) \ .
\end{eqnarray}
In the limit $s\to 0$ the leading term is
\begin{eqnarray}
\label{gleichFA}
\left\{ R (x ,x^\prime )^l \right\}_f^{(c)} &=& v_l \sum_b \left( \left. 
\frac{\partial R (x ,x^\prime )}{\partial \rho_b} \right|_{\overline{\rho}} \right)^l 
\nonumber \\
&=& v_l \sum_b \left( \frac{I_b}{I} \right)^{2l} \ ,
\end{eqnarray}
where we have used Cohn's Theorem Eq.~(\ref{originalCohn}). Upon substitution of 
Eq.~(\ref{gleichFA}) into Eq.~(\ref{defCumulant}) one finds for the noise cumulants
\begin{eqnarray}
\label{finalCumulant}
C_R^{(l)}(x ,x^\prime) = v_l \, M_I^{(l)}(x ,x^\prime) \ , 
\end{eqnarray}
i.e., the noise cumulant $C_R^{(l)}$ is proportional to the $l$th multifractal moment
\begin{eqnarray}
\label{multiMoment}
M_I^{(l)}(x ,x^\prime) = \left\langle \chi (x ,x^\prime) \sum_b 
\left( \frac{I_b}{I} \right)^{2l} \right\rangle_C / \left\langle \chi (x ,x^\prime) 
\right\rangle_C \ , 
\end{eqnarray}
 of the current distribution. 

\subsection{Generating function}
\label{noisyGeneratingFunction}
Our aim is to determine $C_R^{(l)}$. Hence the task is to solve the set of 
Kirchhoff's equations~(\ref{noisyCirquitEquations}) and to perform the averages over the 
diluted lattice configurations and the noise. It can be achieved by employing the 
replica technique\cite{stephen_78}. In order to treat the averages $\langle \cdots 
\rangle_C$ and $\left\{ \cdots \right\}_f$ separately, PHL introduced $D 
\hspace{-1mm}\times \hspace{-1mm}E$-fold replicated voltages,
\begin{eqnarray}
V_x \to \stackrel{\mbox{{\tiny $\leftrightarrow$}}}{V}_x =
\left( 
\begin{array}{ccc}
V_x^{(1,1)} & \cdots & V_x^{(1,D)}\\
\vdots & \ddots & \vdots\\
V_x^{(E,1)} & \cdots & V_x^{(E,D)}
\end{array}
\right) \ .
\end{eqnarray}

Note from the definitions Eq.~(\ref{defCumulant}) and Eq.~(\ref{defOfNoiseCumulant}) 
that one has to treat the two averages independently in the calculation of $C_R^{(n)}$. 
In contrast, for calculating $M_R^{(n)}$ it is not necessary to distinguish between the 
two averages because one could also introduce a composite distribution function
\begin{eqnarray}
f^{\mbox{{\scriptsize comp}}} \left( \sigma \right) = \left( 1 - p \right) \delta \left( 
\sigma \right) + p f \left( \sigma \right)
\end{eqnarray}
and a single, say $D$-fold, replication would be sufficient.

To construct a generating function for the noise cumulants one introduces
\begin{eqnarray}
\psi_{\stackrel{\mbox{{\tiny $\leftrightarrow$}}}{\lambda}}(x) = \exp \left( i 
\stackrel{\mbox{{\tiny $\leftrightarrow$}}}{\lambda} \cdot \stackrel{\mbox{{\tiny 
$\leftrightarrow$}}}{V}_x \right) \ ,
\end{eqnarray}
where $\stackrel{\mbox{{\tiny $\leftrightarrow$}}}{\lambda} \cdot \stackrel{\mbox{{\tiny 
$\leftrightarrow$}}}{V}_x = \sum_{\alpha , \beta =1}^{D,E} \lambda^{(\alpha , \beta )} 
V_x^{(\alpha , \beta)}$ and $\stackrel{\mbox{{\tiny $\leftrightarrow$}}}{\lambda} \neq 
\stackrel{\mbox{{\tiny $\leftrightarrow$}}}{0}$.
The corresponding correlation functions
\begin{eqnarray}
G \left( x, x^\prime ;\stackrel{\mbox{{\tiny $\leftrightarrow$}}}{\lambda} \right) = 
\left\langle \psi_{\stackrel{\mbox{{\tiny 
$\leftrightarrow$}}}{\lambda}}(x)\psi_{-\stackrel{\mbox{{\tiny 
$\leftrightarrow$}}}{\lambda}}(x^\prime) 
\right\rangle_{\mbox{\scriptsize{rep}}}
\end{eqnarray}
are defined as
\begin{eqnarray}
\label{noisyErzeugendeFunktion}
\lefteqn{ G \left( x, x^\prime ;\stackrel{\mbox{{\tiny $\leftrightarrow$}}}{\lambda} 
\right) =
\lim_{D \to 0} \Bigg\langle \Bigg\{ \frac{1}{\prod_{\beta =1}^E Z \left( \left\{ 
\sigma_b^{(\beta )} \right\} , C \right)^D } 
\int \prod_j d \stackrel{\mbox{{\tiny $\leftrightarrow$}}}{V}_j }
\nonumber \\
& & \times \exp \bigg[ -\frac{1}{2} \sum_{\beta =1}^E P \left( \left\{ \vec{V}^{(\beta 
)} 
\right\} , \left\{ \sigma_b^{(\beta )} \right\} , C \right) 
+ \frac{i\omega}{2} \sum_i \stackrel{\mbox{{\tiny $\leftrightarrow$}}}{V}^2_i + i 
\stackrel{\mbox{{\tiny $\leftrightarrow$}}}{\lambda} \cdot \left( \stackrel{\mbox{{\tiny 
$\leftrightarrow$}}}{V}_x  - 
\stackrel{\mbox{{\tiny $\leftrightarrow$}}}{V}_{x^\prime} \right) \bigg] \Bigg\}_f 
\Bigg\rangle_C \ .
\end{eqnarray}
Here $d \stackrel{\mbox{{\tiny $\leftrightarrow$}}}{V}_j = \prod_{\alpha ,\beta=1}^{D,E} 
dV_j^{(\alpha , \beta )}$,
\begin{eqnarray}
P \left( \left\{ \vec{V}^{(\beta )} \right\} , \left\{ \sigma_b^{(\beta )} \right\} , C  
\right) = \sum_{\alpha =1}^{D} \sum_{\langle i,j \rangle} \sigma_{i,j}^{(\beta )} 
\left( V_i^{(\alpha ,\beta)} - V_j^{(\alpha ,\beta )}\right)^2
\end{eqnarray}
with $\vec{V}_x^{(\beta )} = \left( V_x^{(1 , \beta )} , \cdots , V_x^{(D , \beta )} 
\right)$, and $Z$ is the 
normalization
\begin{eqnarray}
\label{noisyNorm}
Z \left( \left\{ \sigma_b^{(\beta )} \right\} , C \right) = \int \prod_{j} dV_{j} \exp 
\left[ -\frac{1}{2} P \left( \left\{ V \right\} , \left\{ \sigma_b^{(\beta )} \right\} , 
C 
\right) + \frac{i\omega}{2} \sum_i V^2_i \right] \ .
\end{eqnarray}
Note that we have introduced an additional power term $\frac{i\omega}{2} \sum_i V^2_i$. 
This is necessary to give the integrals in Eqs.~(\ref{noisyErzeugendeFunktion}) and 
(\ref{noisyNorm}) a well defined meaning. Without this term the integrands depend only 
on voltage differences and the integrals are divergent. Physically the new term 
corresponds to grounding each lattice site by a capacitor of unit capacity. The original 
situation may be restored by taking the limit of vanishing frequency, $\omega \to 0$.

The integrations in Eq.~(\ref{noisyErzeugendeFunktion}) can be carried out by employing 
the saddle point method. Since the integrations are Gaussian the saddle point method is 
exact in this case. The saddle point equation is identical to the variation principle 
stated in Eq.~(\ref{noisyVariationPrinciple1}). Thus the maximum of the integrand is 
determined by the solution of Kirchhoff's equations (\ref{noisyCirquitEquations}) and 
\begin{eqnarray}
\label{noisyGenFkt}
G \left( x, x^\prime ;\stackrel{\mbox{{\tiny $\leftrightarrow$}}}{\lambda} \right) = 
\left\langle \prod_{\beta =1}^E \left\{ \exp \left[ - 
\frac{\vec{\lambda}^{(\beta )2}}{2} R^{(\beta )} \left( x,x^\prime 
\right) \right] \right\}_f \right\rangle_C \ .
\end{eqnarray}
 The right hand side of Eq.~(\ref{noisyGenFkt}) may be expanded in term of the cumulants 
defined in Eq.~(\ref{defOfNoiseCumulant}). This gives
\begin{eqnarray}
\label{cumulantGenFkt}
G \left( x, x^\prime ;\stackrel{\mbox{{\tiny $\leftrightarrow$}}}{\lambda} \right) = 
\left\langle  \exp \left[ \sum_{l=1}^\infty \frac{(-1/2)^l}{l!} K_l \left( 
\stackrel{\mbox{{\tiny $\leftrightarrow$}}}{\lambda} \right) \left\{ R 
\left( x,x^\prime \right)^l \right\}_f^{(c)} \right]  \right\rangle_C \ ,
\end{eqnarray}
where $K_l$ is defined by
\begin{eqnarray}
\label{defOfK_l}
K_l \left( \stackrel{\mbox{{\tiny $\leftrightarrow$}}}{\lambda} \right) = \sum_{\beta 
=1}^E \left[ \sum_{\alpha =1}^D \left( \lambda^{(\alpha ,\beta )} \right)^2 \right]^l \ 
.
\end{eqnarray}
We learn that the correlation function $G$ can be exploited as a generating function for 
the noise cumulants via 
\begin{eqnarray}
\label{exploitGenFkt}
\left\langle \chi ( x, x^\prime) \right\rangle_C C_R^{(n)} \left( x, 
x^\prime \right) = \frac{\partial}{\partial \, \frac{(-1/2)^n}{n!} K_n \left( 
\stackrel{\mbox{{\tiny $\leftrightarrow$}}}{\lambda} \right)} G \left( x, x^\prime 
;\stackrel{\mbox{{\tiny $\leftrightarrow$}}}{\lambda} \right) 
\bigg|_{\stackrel{\mbox{{\tiny $\leftrightarrow$}}}{\lambda} = \stackrel{\mbox{{\tiny 
$\leftrightarrow$}}}{0}} \ .
\end{eqnarray}
Note that $M_R^{(1)} = C_R^{(1)}$.

\subsection{Field theoretic Hamiltonian}
\label{noisyFieldTheoreticHamiltonian}
Since infinite voltage drops between different clusters may occur, it is not guaranteed 
that $Z$ stays finite, i.e., the limit $\lim_{D \to 0}Z^{DE}$ is not well defined. 
Moreover, $\stackrel{\mbox{{\tiny $\leftrightarrow$}}}{\lambda} = \stackrel{\mbox{{\tiny 
$\leftrightarrow$}}}{0}$ has to be excluded properly. Both problems can be handled by 
resorting to a lattice regularization of the integrals in 
Eqs.~(\ref{noisyErzeugendeFunktion}) and (\ref{noisyNorm}). One switches to voltage 
variables $\stackrel{\mbox{{\tiny $\leftrightarrow$}}}{\theta}= \Delta \theta 
\stackrel{\mbox{{\tiny $\leftrightarrow$}}}{k}$ taking 
discrete values on a $D \hspace{-1mm}\times \hspace{-1mm}E$-dimensional torus, i.e., 
$\stackrel{\mbox{{\tiny $\leftrightarrow$}}}{k}$ is chosen to be an 
$D \hspace{-1mm}\times \hspace{-1mm}E$-dimensional integer with $-M < k^{(\alpha ,\beta 
)} \leq M$ and $k^{(\alpha ,\beta)}=k^{(\alpha ,\beta )} \mbox{mod} (2M)$. $\Delta 
\theta = 
\theta_M /M$ is the gap between successive voltages and $\theta_M$ is the voltage 
cutoff. 
The continuum may be restored by taking $\theta_M \to \infty$ and $\Delta \theta \to 0$. 
By 
setting $\theta_M = \theta_0 M$, $M=m^2$, and, respectively, $\Delta \theta = \theta_0 
/m$, 
the two limits can be taken simultaneously via $m \to \infty$. Since the voltage and 
current variables are conjugated $\stackrel{\mbox{{\tiny $\leftrightarrow$}}}{\lambda}$ 
is 
affected by the discretization as well:
\begin{eqnarray}
\stackrel{\mbox{{\tiny $\leftrightarrow$}}}{\lambda} = \Delta \lambda 
\stackrel{\mbox{{\tiny $\leftrightarrow$}}}{l} \ , \ \Delta \lambda \hspace{1mm} \Delta 
\theta = \pi / M \ ,
\end{eqnarray}
where $\stackrel{\mbox{{\tiny $\leftrightarrow$}}}{l}$ is a $D \hspace{-1mm}\times 
\hspace{-1mm}E$-dimensional integer taking the same values as $\stackrel{\mbox{{\tiny 
$\leftrightarrow$}}}{k}$. This choice guarantees that the completeness and 
orthogonality relations
\begin{mathletters}
\label{noisyComplete}
\begin{eqnarray}
\frac{1}{(2M)^{DE}} \sum_{\stackrel{\mbox{\tiny $\leftrightarrow$}}{\theta}} \exp \left( 
i \stackrel{\mbox{{\tiny $\leftrightarrow$}}}{\lambda} \cdot \stackrel{\mbox{{\tiny 
$\leftrightarrow$}}}{\theta} \right) = \delta_{\stackrel{\mbox{{\tiny 
$\leftrightarrow$}}}{\lambda} ,\stackrel{\mbox{{\tiny $\leftrightarrow$}}}{0} 
\hspace{0.15em}\mbox{\scriptsize{mod}}(2M \Delta \lambda) }
\end{eqnarray}
and
\begin{eqnarray}
\frac{1}{(2M)^{DE}} \sum_{\stackrel{\mbox{{\tiny $\leftrightarrow$}}}{\lambda}} \exp 
\left( i \stackrel{\mbox{{\tiny $\leftrightarrow$}}}{\lambda} \cdot 
\stackrel{\mbox{{\tiny 
$\leftrightarrow$}}}{\theta} \right) = \delta_{\stackrel{\mbox{{\tiny 
$\leftrightarrow$}}}{\theta} ,\stackrel{\mbox{{\tiny $\leftrightarrow$}}}{0} 
\hspace{0.15em}\mbox{\scriptsize{mod}}(2M \Delta \theta)}
\end{eqnarray}
\end{mathletters}
do hold. Equation~(\ref{noisyComplete}) provides us with a Fourier transform in replica 
space. In this discrete picture there are $(2M)^{DE} -1$ independent state variables per 
lattice site. Upon Fourier transformation one introduces the Potts 
spins\cite{Zia_Wallace_75} 
\begin{eqnarray}
\Phi_{\stackrel{\mbox{\tiny $\leftrightarrow$}}{\theta}} \left( x \right) = (2M)^{-DE} 
\sum_{\stackrel{\mbox{\tiny $\leftrightarrow$}}{\lambda} \neq 
\stackrel{\mbox{\tiny $\leftrightarrow$}}{0}} \exp \left( i \stackrel{\mbox{\tiny 
$\leftrightarrow$}}{\lambda} \cdot \stackrel{\mbox{\tiny $\leftrightarrow$}}{\theta} 
\right) \psi_{\stackrel{\mbox{\tiny $\leftrightarrow$}}{\lambda}} (x) = 
\delta_{\stackrel{\mbox{\tiny $\leftrightarrow$}}{\theta}, \stackrel{\mbox{\tiny 
$\leftrightarrow$}}{\theta}_{x}} - (2M)^{-DE} 
\end{eqnarray}
subject to the condition $\sum_{\stackrel{\mbox{\tiny $\leftrightarrow$}}{\theta }} 
\Phi_{\stackrel{\mbox{\tiny $\leftrightarrow$}}{\theta}} \left( x \right) = 0$.

Now we revisit Eq.~(\ref{noisyErzeugendeFunktion}). Carrying out the average over the 
diluted lattice configurations and the noise provides us with the weight $\exp 
(-H_{\mbox{\scriptsize{rep}}})$ of the average $\langle \cdots 
\rangle_{\mbox{\scriptsize{rep}}}$,
\begin{eqnarray}
H_{\mbox{\scriptsize{rep}}} &=&  - \ln \left\langle \left\{ \exp \left[ - \frac{1}{2} P
\left( \left\{ \stackrel{\mbox{\tiny $\leftrightarrow$}}{\theta} \right\} \right) + \frac{i\omega}{2} 
\sum_i \stackrel{\mbox{\tiny $\leftrightarrow$}}{\theta}_{i}^2 \right] \right\}_f 
\right\rangle_C 
\nonumber \\
&=& - \sum_{\langle i ,j \rangle} \ln \left\langle \prod_{\beta =1}^E \left\{ \exp 
\left[ - 
\frac{1}{2} \sigma_{i ,j}^{(\beta )} \left( \vec{\theta}_{i}^{(\beta )} - 
\vec{\theta}_{j}^{(\beta )} \right)^2 \right] \right\}_f 
\right\rangle_C - \frac{i\omega}{2} \sum_i \stackrel{\mbox{\tiny 
$\leftrightarrow$}}{\theta}_{i}^2 \ .
\end{eqnarray}
By dropping a constant term $N_B \ln (1-p)$, with $N_B$ being the number of 
bonds in the undiluted lattice, one obtains
\begin{eqnarray}
H_{\mbox{\scriptsize{rep}}} &=& - \sum_{\langle i ,j \rangle} K \left( 
\stackrel{\mbox{\tiny $\leftrightarrow$}}{\theta}_{i} - \stackrel{\mbox{\tiny 
$\leftrightarrow$}}{\theta}_{j} \right) - \sum_i h \left( \stackrel{\mbox{\tiny 
$\leftrightarrow$}}{\theta}_{i} \right)
\nonumber \\
&=& - \sum_{\langle i ,j \rangle} \sum_{\stackrel{\mbox{\tiny 
$\leftrightarrow$}}{\theta},\stackrel{\mbox{\tiny $\leftrightarrow$}}{\theta}^\prime} K 
\left( \stackrel{\mbox{\tiny $\leftrightarrow$}}{\theta} - \stackrel{\mbox{\tiny 
$\leftrightarrow$}}{\theta}^\prime \right) \Phi_{\stackrel{\mbox{\tiny 
$\leftrightarrow$}}{\theta}} \left( i \right) \Phi_{\stackrel{\mbox{\tiny 
$\leftrightarrow$}}{\theta}^\prime} \left( j \right) - \sum_i 
\sum_{\stackrel{\mbox{\tiny 
$\leftrightarrow$}}{\theta}} h \left( \stackrel{\mbox{\tiny $\leftrightarrow$}}{\theta} 
\right) \Phi_{\vec{\theta}} \left( i \right) \ ,
\end{eqnarray}
where
\begin{eqnarray}
h \left( \stackrel{\mbox{\tiny $\leftrightarrow$}}{\theta} \right) = \frac{i\omega}{2} 
\sum_i \stackrel{\mbox{\tiny $\leftrightarrow$}}{\theta}_{i}^2 
\end{eqnarray}
and
\begin{eqnarray}
K \left( \stackrel{\mbox{\tiny $\leftrightarrow$}}{\theta} \right) &=& \ln \left\{ 1 + 
\frac{p}{1-p} \prod_{\beta =1}^E \left\{ \exp \left[ - 
\frac{1}{2} \sigma^{(\beta )} \sum_{\alpha =1}^D \left( \theta^{(\alpha ,\beta)} 
\right)^2 
\right] 
\right\}_f \right\}
\nonumber \\
&=& \ln \left\{ 1 + \frac{p}{1-p} \exp \left[ \sum_{l=1}^\infty \frac{(-1/2)^l}{l!} 
\Delta^{(l)} K_l \left( \stackrel{\mbox{\tiny $\leftrightarrow$}}{\theta} \right) 
\right] 
\right\} \ .
\end{eqnarray}

In the limit of perfect transport, $s \to 0$, $K \left( \stackrel{\mbox{\tiny 
$\leftrightarrow$}}{\theta} \right)$ goes to its local limit $K \left( 
\stackrel{\mbox{\tiny $\leftrightarrow$}}{\theta} \right) = K 
\delta_{\stackrel{\mbox{\tiny 
$\leftrightarrow$}}{\theta}, \stackrel{\mbox{\tiny $\leftrightarrow$}}{0}}$, with $K$ 
being 
a positive constant. The interaction part of the Hamiltonian reduces to
\begin{eqnarray}
\label{interActionHamil}
H^{\mbox{\scriptsize{int}}}_{\mbox{\scriptsize{rep}}} = - K \sum_{\langle i ,j \rangle} 
\sum_{\vec{\theta}} \Phi_{\stackrel{\mbox{\tiny $\leftrightarrow$}}{\theta}} \left( i 
\right) \Phi_{\stackrel{\mbox{\tiny $\leftrightarrow$}}{\theta}} \left( j \right) \ .
\end{eqnarray}
This represents nothing more than the $\left( 2M \right)^{DE}$ states Potts model which 
is invariant against all $\left( 2M \right)^{DE} !$ permutations of the Potts spins 
$\Phi_{\stackrel{\mbox{\tiny $\leftrightarrow$}}{\theta}}$.

In the case of imperfect transport this $S_{\left( 2M \right)^{DE}}$ symmetry is lost. 
For finite $\overline{\sigma}$ and $\Delta^{(n)} =0$, $K \left( \stackrel{\mbox{\tiny 
$\leftrightarrow$}}{\theta} \right)$ is an exponentially decreasing function in replica 
space with a decay rate proportional to $\overline{\sigma}^{-1}$. Then, for large 
$\overline{\sigma}$, the Hamiltonian $H_{\mbox{\scriptsize{rep}}}$ describes a 
translationally and rotationally invariant short range interaction of Potts spins in real 
and replica space with an external one site potential $ h \left( \stackrel{\mbox{\tiny 
$\leftrightarrow$}}{\theta} \right)$.

Admitting fluctuations of the resistances, $\Delta^{(n)} > 0$, results in breaking the 
rotational $O (DE)$ replica space symmetry of the interaction part of the Hamiltonian. 
The Fourier transform of $K \left( \stackrel{\mbox{\tiny $\leftrightarrow$}}{\theta} 
\right)$,
\begin{eqnarray}
\widetilde{K} \left( \stackrel{\mbox{\tiny $\leftrightarrow$}}{\lambda} \right) = 
\frac{1}{(2M)^{DE}} \sum_{\stackrel{\mbox{\tiny $\leftrightarrow$}}{\theta}} 
\exp \left( -i \stackrel{\mbox{\tiny $\leftrightarrow$}}{\lambda} \cdot 
\stackrel{\mbox{\tiny $\leftrightarrow$}}{\theta} \right) K \left( \stackrel{\mbox{\tiny 
$\leftrightarrow$}}{\theta} 
\right)
\end{eqnarray}
is expediently evaluated by switching back to continuous voltages,
\begin{eqnarray}
\widetilde{K} \left( \stackrel{\mbox{\tiny $\leftrightarrow$}}{\lambda} \right) = 
\int_{-\infty}^\infty d \stackrel{\mbox{\tiny $\leftrightarrow$}}{\theta} \ \exp \left( 
-i 
\stackrel{\mbox{\tiny $\leftrightarrow$}}{\lambda} \cdot 
\stackrel{\mbox{\tiny $\leftrightarrow$}}{\theta} \right) \ln \left\{ 1 + \frac{p}{1-p} 
\exp \left[ \sum_{l=1}^\infty \frac{(-1/2)^l}{l!} \Delta^{(l)} K_l \left( 
\stackrel{\mbox{\tiny $\leftrightarrow$}}{\theta} \right) \right] \right\} \ , 
\end{eqnarray}
where we have dropped a factor $(2\theta_M )^{-DE}$. Taylor expansion of the logarithm 
yields a series of terms of the form
\begin{eqnarray}
\label{s1} 
\int_{-\infty}^\infty d \stackrel{\mbox{\tiny $\leftrightarrow$}}{\theta} \ \exp \left[ 
-i 
\stackrel{\mbox{\tiny $\leftrightarrow$}}{\lambda} \cdot 
\stackrel{\mbox{\tiny $\leftrightarrow$}}{\theta} - a \overline{\sigma} 
\stackrel{\mbox{\tiny $\leftrightarrow$}}{\theta}^2 - \sum_{l=2} b_l \left( 
\overline{\sigma} s \right)^l K_l \left( \stackrel{\mbox{\tiny 
$\leftrightarrow$}}{\theta} 
\right) \right]  \ , 
\end{eqnarray}
where $a$ the $b_l$ are constants of order ${\sl O} \left( s^0 \right)$. In addition to 
the expansion of the logarithm we expand in a power series in $s$, 
\begin{eqnarray}
\label{s2} 
\mbox{(\ref{s1})} = \int_{-\infty}^\infty d \stackrel{\mbox{\tiny 
$\leftrightarrow$}}{\theta} \ \exp \left[ -i \stackrel{\mbox{\tiny 
$\leftrightarrow$}}{\lambda} \cdot 
\stackrel{\mbox{\tiny $\leftrightarrow$}}{\theta} - a \overline{\sigma} 
\stackrel{\mbox{\tiny $\leftrightarrow$}}{\theta}^2 \right] \left\{ 1 + 
\sum_{l=2}^\infty 
\left( \overline{\sigma} s \right)^l P_l \left( \stackrel{\mbox{\tiny 
$\leftrightarrow$}}{\theta} \right) \right\} \ . 
\end{eqnarray}
Here the $P_l$ are homogeneous polynomials of order $2l$ in $\stackrel{\mbox{\tiny 
$\leftrightarrow$}}{\lambda}$ which are a sums of terms proportinal to
\begin{eqnarray}
\prod_{i \geq 2} K_i \left( \stackrel{\mbox{\tiny $\leftrightarrow$}}{\theta} 
\right)^{l_i} 
\end{eqnarray} 
such that $\sum_i i l_i = l$. Completing squares in the exponential in Eq.~(\ref{s2}) 
gives
\begin{eqnarray}
\label{s3} 
\mbox{(\ref{s2})} &=& \exp \left[ - \frac{\stackrel{\mbox{\tiny 
$\leftrightarrow$}}{\lambda}^2}{4 a \overline{\sigma}} \right] \int_{-\infty}^\infty d 
\stackrel{\mbox{\tiny $\leftrightarrow$}}{\theta} \ \exp \left[ - a \overline{\sigma} 
\stackrel{\mbox{\tiny $\leftrightarrow$}}{\theta}^2 \right] \left\{ 1 + 
\sum_{l=2}^\infty 
\left( \overline{\sigma} s \right)^l P_l \left( \stackrel{\mbox{\tiny 
$\leftrightarrow$}}{\theta} - i \frac{\stackrel{\mbox{\tiny 
$\leftrightarrow$}}{\lambda}}{2 
a \overline{\sigma}} \right) \right\} 
\nonumber \\
&=& \exp \left[ - \frac{\stackrel{\mbox{\tiny $\leftrightarrow$}}{\lambda}^2}{4 a 
\overline{\sigma}} \right] \left\{ 1 + \sum_{l=2}^\infty \left( \overline{\sigma} s 
\right)^l \bigg[ P_l \left( \frac{\stackrel{\mbox{\tiny $\leftrightarrow$}}{\lambda}}{ 
\overline{\sigma}} \right) + \cdots + \overline{\sigma}^{(-r)} P_{l-r} \left( 
\frac{\stackrel{\mbox{\tiny $\leftrightarrow$}}{\lambda}}{ \overline{\sigma}} \right) + 
\cdots \bigg] \right\} \ , 
\end{eqnarray}
where we have omitted multiplicative factors decorating the $P_l$. Due to the homogenity 
of the $P_l$, Eq.~(\ref{s3}) can be rearranged as
\begin{eqnarray}
\label{s4} 
\mbox{(\ref{s3})} &=& \exp \left[ - \frac{\stackrel{\mbox{\tiny 
$\leftrightarrow$}}{\lambda}^2}{4 a \overline{\sigma}} \right] \left\{ 1 + 
\sum_{l=2}^\infty  s^l \bigg[ \overline{\sigma}^{-l} P_l \left( \stackrel{\mbox{\tiny 
$\leftrightarrow$}}{\lambda} \right) + \cdots + \overline{\sigma}^{-(l-r)}P_{l-r} \left( 
\stackrel{\mbox{\tiny $\leftrightarrow$}}{\lambda} \right) + \cdots \bigg] \right\} 
\nonumber \\
&=& \exp \left[ - \frac{\stackrel{\mbox{\tiny $\leftrightarrow$}}{\lambda}^2}{4 a 
\overline{\sigma}} \right] \left\{ 1 + \sum_{l^\prime =1}^\infty  \left( 
\frac{s}{\overline{\sigma}} \right)^{l^\prime} \bigg[ 1 + {\sl O} \left( s \right) 
\bigg] 
P_{l^\prime} \left( \stackrel{\mbox{\tiny $\leftrightarrow$}}{\lambda} \right) \right\} 
\ , 
\end{eqnarray}
up to multiplicative factors. By keeping only the leading contributions, one finds that 
$\widetilde{K} \left( \stackrel{\mbox{\tiny $\leftrightarrow$}}{\lambda} \right)$ can be 
expanded as 
\begin{eqnarray}
\widetilde{K} \left( \stackrel{\mbox{\tiny $\leftrightarrow$}}{\lambda} \right) = \tau + 
\sum_{p=1}^\infty w_p \stackrel{\mbox{\tiny $\leftrightarrow$}}{\lambda}^{2p} + 
\sum_{P_l} v_{P_l} P_l \left( \stackrel{\mbox{\tiny 
$\leftrightarrow$}}{\lambda} 
\right) \ ,
\end{eqnarray}
with $\tau$, $w_p \sim \overline{\sigma}^{-p}$, and $v_{P_l} \sim \Delta^{(l)} / 
\overline{\sigma}^{2l}$ being expansion coefficients.

It is known that the terms $w_p \stackrel{\mbox{\tiny $\leftrightarrow$}}{\lambda}^{2p}$ 
are irrelevant in the renormalization group sense for $p \geq 2$ (see, 
e.g.\cite{stenull_janssen_oerding_99}). From Sec.~\ref{relevanceOfTheNoiseTerms} can be 
inferred that the $v_{P_l} P_l \left( \stackrel{\mbox{\tiny $\leftrightarrow$}}{\lambda} 
\right)$ are irrelevant as well. However, the terms proportional to $K_l \left( 
\stackrel{\mbox{\tiny $\leftrightarrow$}}{\lambda} \right)$ are indispenseble in 
studying the noise cumulants; they are dangerously irrelevant. Therefore, we restrict the 
expansion of $\widetilde{K} \left( \stackrel{\mbox{\tiny $\leftrightarrow$}}{\lambda} 
\right)$ to
\begin{eqnarray}
\label{noisyTaylorExp}
\widetilde{K} \left( \stackrel{\mbox{\tiny $\leftrightarrow$}}{\lambda} \right) = \tau + 
w \stackrel{\mbox{\tiny $\leftrightarrow$}}{\lambda}^2 + \sum_{l=2}^{\infty} v_l K_l 
\left( 
\stackrel{\mbox{\tiny $\leftrightarrow$}}{\lambda} \right) \ ,
\end{eqnarray}
with $w = w_1$, and $v_{l} = v_{K_l}$. Nevertheless the neglected terms will regain some 
importance later on since they are required for the renormalization of the $v_l$. 

The $K_l$ are homogeneous polynomials of order $2l$. For $l\geq 2$ they are possessing a 
$S \left[ O \left (D \right)^E \right]$ symmetry. Thus, allowing for $\Delta^{(n)} >0$ 
results in loosing the rotational $O \left( DE\right)$ in favor of the $S \left[ O \left 
(D \right)^E \right]$ symmetry.

It is worth pointing out that $v_{l}/w^{l} \sim \Delta^{(l)}/\overline{\sigma}^l \sim 
s^l$, i.e., the condition $s \to 0$ translates into $v_l \ll w^l$. Consequently one has 
to take the limit $v_l \to 0$ before the limit $w \to 0$ in calculating the exponents 
associated with the $v_l$.

We proceed with the usual coarse graining step and replace the Potts spins 
$\Phi_{\stackrel{\mbox{{\tiny $\leftrightarrow$}}}{\theta}} \left( x \right)$ by order 
parameter fields $\varphi \left( {\rm{\bf x}} ,\stackrel{\mbox{{\tiny 
$\leftrightarrow$}}}{\theta} \right)$ which inherit the constraint 
$\sum_{\stackrel{\mbox{{\tiny $\leftrightarrow$}}}{\theta}} \varphi \left( {\rm{\bf x}} 
,\stackrel{\mbox{{\tiny $\leftrightarrow$}}}{\theta} \right) = 0$. We model the 
corresponding field theoretic Hamiltonian $\mathcal{H}$ in the spirit 
of Landau as a mesoscopic free energy from local monomials of the order parameter field 
and its gradients in real and replica space. The gradient expansion is justified since 
the interaction is short ranged in both spaces. Purely local terms in replica space have 
to respect the full $S_{\left( 2M \right)^{DE}}$ Potts symmetry. After these remarks we 
write down the Landau-Ginzburg-Wilson type Hamiltonian
\begin{eqnarray}
\label{noisyFinalHamil}
{\mathcal{H}} = \int d^dx \sum_{\stackrel{\mbox{{\tiny 
$\leftrightarrow$}}}{\theta}} 
\Bigg\{  \frac{1}{2} \varphi \left( {\rm{\bf x}} , 
\stackrel{\mbox{{\tiny $\leftrightarrow$}}}{\theta} \right) K \left( \Delta ,
\nabla_{\stackrel{\mbox{{\tiny $\leftrightarrow$}}}{\theta}} \right) \varphi \left( 
{\rm{\bf x}} , \stackrel{\mbox{{\tiny $\leftrightarrow$}}}{\theta} \right)
+ \frac{g}{6}\varphi \left( {\rm{\bf x}} , \stackrel{\mbox{{\tiny 
$\leftrightarrow$}}}{\theta} \right)^3 + \frac{i \omega}{2} \stackrel{\mbox{{\tiny 
$\leftrightarrow$}}}{\theta}^2 \varphi \left( {\rm{\bf x}} , \stackrel{\mbox{{\tiny 
$\leftrightarrow$}}}{\theta} \right) \Bigg\} \ ,
\end{eqnarray}
where
\begin{eqnarray}
K \left( \Delta ,\nabla_{\stackrel{\mbox{{\tiny $\leftrightarrow$}}}{\theta}} \right) = 
\tau + \Delta + w \sum_{\alpha , \beta =1}^{D,E} \frac{- \partial^2}{\left( \partial 
\theta^{(\alpha ,\beta 
)} \right)^2 } + \sum_{l=2}^\infty v_l \sum_{\beta=1}^E \left[ \sum_{\alpha =1}^D   
\frac{- \partial^2}{\left( \partial \theta^{(\alpha ,\beta )} \right)^2 } \right]^{l} \ 
.
\end{eqnarray}
In Eq.~(\ref{noisyFinalHamil}) we have neglected terms of order $\varphi^4$ or higher 
which are irrelevant in the renormalization group sense. $\tau$, $w$ and $v_l$ are now 
coarse grained analogues of the original coefficients appearing in 
Eq.~(\ref{noisyTaylorExp}). Note again that $\mathcal{H}$ reduces to the 
usual $(2M)^{DE}$ states Potts model Hamiltonian by setting $v_l =0$ and $w=0$ as one 
retrieves purely geometrical percolation in the limit of vanishing $v_l$ and $w$. 

\subsection{Relevance of the noise terms}
\label{relevanceOfTheNoiseTerms}
Irrelevant variables that cannot be taken to zero because the quantity one is looking at either vanishes or diverges in this limit have been given the name {\sl dangerously irrelevant variables} by Fisher\cite{fisher_74}. Later on this notion was introduced into field theory by Amit and Peliti\cite{amit_peliti_82}. A characteristic feature of dangerously irrelevant variables is that corrections due to them determine the asymptotic behavior of quantities with the above property, so that their effect is felt arbitrarily close to the transition\cite{diehl_86}. In this section we show that the $v_l$ are dangerously irrelevant. They are irrelevant on dimensional grounds, i.e., they are associated with a negative naive dimension. However, we cannot simply take the $v_l$ to zero by appealing to their irrelevance, because the amplitudes of the noise cumulants vanish in this limit.

In the remainder of this article we focus on vanishing frequency, $\omega =0$. Let $P$ 
denote the set of parameters $\left\{ \tau ,w , v_l \right\}$. We introduce a scaling 
factor $b$ for the voltage variable: $\stackrel{\mbox{{\tiny $\leftrightarrow$}}}{\theta} 
\to b \stackrel{\mbox{{\tiny $\leftrightarrow$}}}{\theta}$. By substitution of 
$\varphi \left( {\rm{\bf x}} , \stackrel{\mbox{{\tiny $\leftrightarrow$}}}{\theta} 
\right) = \varphi^\prime \left( {\rm{\bf x}} , b\stackrel{\mbox{{\tiny 
$\leftrightarrow$}}}{\theta} \right)$ the Hamiltonian turns into
\begin{eqnarray}
\label{noisyScaling1}
{\mathcal{H}} \left[ \varphi^\prime \left( {\rm{\bf x}} , b 
\stackrel{\mbox{{\tiny $\leftrightarrow$}}}{\theta} \right) , P \right] = \int d^dx 
\sum_{\stackrel{\mbox{{\tiny $\leftrightarrow$}}}{\theta}} \left\{ \frac{1}{2} 
\varphi^\prime \left( {\rm{\bf x}} , b \stackrel{\mbox{{\tiny 
$\leftrightarrow$}}}{\theta} \right) K \left( \Delta ,
\nabla_{\stackrel{\mbox{{\tiny $\leftrightarrow$}}}{\theta}} \right) 
\varphi^\prime \left( {\rm{\bf x}} , b \stackrel{\mbox{{\tiny 
$\leftrightarrow$}}}{\theta} 
\right) + \frac{g}{6}\varphi^{\prime } \left( {\rm{\bf x}} , b \stackrel{\mbox{{\tiny 
$\leftrightarrow$}}}{\theta} \right)^3 \right\} \ .
\end{eqnarray}
Renaming the scaled voltage variables $\stackrel{\mbox{{\tiny 
$\leftrightarrow$}}}{\theta}^\prime = b \stackrel{\mbox{{\tiny 
$\leftrightarrow$}}}{\theta}$ yields
\begin{eqnarray}
\label{noisyScaling2}
{\mathcal{H}} \left[ \varphi^\prime \left( {\rm{\bf x}} ,  
\stackrel{\mbox{{\tiny 
$\leftrightarrow$}}}{\theta}^\prime \right) , P \right] = \int d^dx 
\sum_{\stackrel{\mbox{{\tiny $\leftrightarrow$}}}{\theta}^\prime} \left\{ \frac{1}{2} 
\varphi^\prime \left( {\rm{\bf x}} , \stackrel{\mbox{{\tiny 
$\leftrightarrow$}}}{\theta}^\prime \right) K \left( \Delta ,
b \nabla_{\stackrel{\mbox{{\tiny $\leftrightarrow$}}}{\theta}^\prime} \right) 
\varphi^\prime \left( {\rm{\bf x}} , \stackrel{\mbox{{\tiny 
$\leftrightarrow$}}}{\theta}^\prime \right) + \frac{g}{6}\varphi^{\prime } \left( 
{\rm{\bf x}} , \stackrel{\mbox{{\tiny $\leftrightarrow$}}}{\theta}^\prime 
\right)^3 \right\} \ .
\end{eqnarray}
Obviously the voltage cutoff is affected by the scaling as well: $\theta_M \to b 
\theta_M$. However, if the limits are taken in the appropriate order, namely $D \to 0$ 
and then $m \to \infty$, the dependence of the theory on the cutoff drops out. Thus, we 
can identify $\stackrel{\mbox{{\tiny $\leftrightarrow$}}}{\theta}^\prime$ and 
$\stackrel{\mbox{{\tiny $\leftrightarrow$}}}{\theta}$ and hence 
\begin{eqnarray}
\label{noisyRelForH}
{\mathcal{H}} \left[ \varphi \left( {\rm{\bf x}} , b \stackrel{\mbox{{\tiny 
$\leftrightarrow$}}}{\theta} \right) , P \right] = {\mathcal{H}} \left[ \varphi 
\left( {\rm{\bf x}} , \stackrel{\mbox{{\tiny $\leftrightarrow$}}}{\theta} \right) , 
P^\prime \right] \ ,
\end{eqnarray}
where $P^\prime = \left\{ \tau , b^2 w , b^{2l} v_l \right\}$.

Now consider correlation functions 
\begin{eqnarray}
\label{correl}
G_N \left( \left\{ {\rm{\bf x}} ,\stackrel{\mbox{{\tiny $\leftrightarrow$}}}{\theta} 
\right\} ; \tau , w, \left\{ v_l \right\}  \right) = \int {\mathcal{D}} 
\varphi \ \varphi \left( {\rm{\bf x}}_1 , \stackrel{\mbox{{\tiny 
$\leftrightarrow$}}}{\theta}_1 \right) \cdots \varphi \left( {\rm{\bf x}}_N , 
\stackrel{\mbox{{\tiny $\leftrightarrow$}}}{\theta}_N \right)  \exp 
\left( - \mathcal{H} \left[ \varphi \left( {\rm{\bf x}} , 
\stackrel{\mbox{{\tiny $\leftrightarrow$}}}{\theta} \right) , P \right] \right) \ ,
\end{eqnarray}
where ${\mathcal{D}} \varphi$ indicates an integration over the set of 
variables 
$\left\{ \varphi \left( {\rm{\bf x}} , 
\stackrel{\mbox{{\tiny $\leftrightarrow$}}}{\theta} \right) \right\}$ for all ${\rm{\bf 
x}}$ and $\stackrel{\mbox{{\tiny $\leftrightarrow$}}}{\theta}$. 
Equation~(\ref{noisyRelForH}) implies 
\begin{eqnarray}
G_N \left( \left\{ {\rm{\bf x}} ,\stackrel{\mbox{{\tiny $\leftrightarrow$}}}{\theta} 
\right\} ; \tau , w, \left\{ v_l \right\} \right) = G_N \left( \left\{ {\rm{\bf x}} , b 
\stackrel{\mbox{{\tiny $\leftrightarrow$}}}{\theta} \right\} ; \tau , b^2 w, \left\{ 
b^{2l} v_l \right\} \right) \ .
\end{eqnarray}
The two-point correlation function $G_2$ is the Fourier transform of $ \left\langle 
\psi_{\stackrel{\mbox{{\tiny $\leftrightarrow$}}}{\lambda}}({\rm{\bf x}}) 
\psi_{-\stackrel{\mbox{{\tiny $\leftrightarrow$}}}{\lambda}}({\rm{\bf x}}) 
\right\rangle_{\mathcal{H}}$. We deduce from Eq.~(\ref{cumulantGenFkt}) that
\begin{eqnarray}
K_l \left( \stackrel{\mbox{{\tiny $\leftrightarrow$}}}{\lambda} \right) C_R^{(l)} \left( 
\left( {\rm{\bf x}}, {\rm{\bf x}}^\prime \right) ; \tau , w, \left\{ v_k \right\} 
\right) 
= b^{-2l} K_l \left( \stackrel{\mbox{{\tiny $\leftrightarrow$}}}{\lambda} \right) 
C_R^{(l)} \left( \left( {\rm{\bf x}}, {\rm{\bf x}}^\prime \right) ; \tau , b^2 w, 
\left\{ b^{2k} v_k \right\} \right) \ .
\end{eqnarray}
We are free to choose $b^2 = w^{-1}$. This gives
\begin{eqnarray}
\label{cumulantScaling}
C_R^{(l)} \left( \left( {\rm{\bf x}}, {\rm{\bf x}}^\prime \right) ; \tau , w, \left\{ 
v_k 
\right\} \right) = w^l f_l \left( \left( {\rm{\bf x}}, {\rm{\bf x}}^\prime \right) ; 
\tau 
, \left\{ \frac{v_k}{w^k} \right\} \right) \ ,
\end{eqnarray}
where $f_l$ is a scaling function. We learn from Eq.~(\ref{cumulantScaling}) that the 
coupling constants $v_k$ appear only as $v_k / w^k$. Dimensional analysis of the 
Hamiltonian shows that $w \stackrel{\mbox{{\tiny $\leftrightarrow$}}}{\lambda}^2 \sim 
\mu^2$ and  $v_k K_k \left( \stackrel{\mbox{{\tiny $\leftrightarrow$}}}{\lambda} \right) 
\sim \mu^2$, where $\mu$ is an inverse length scale, i.e., $w \stackrel{\mbox{{\tiny 
$\leftrightarrow$}}}{\lambda}^2$ and $v_k K_k \left( \stackrel{\mbox{{\tiny 
$\leftrightarrow$}}}{\lambda} \right)$ have a naive dimension 2. Thus $v_k / w^k \sim 
\mu^{2-2k}$ and hence the $v_k / w^k$ have a negative naive dimension. 
This leads to the conclusion that the $v_{k}$ are irrelevant couplings. 

Though irrelevant, one must not set $v_{l} =0$ in calculating the noise exponents. 
In order to see this we expand the scaling function $f_l$ in 
Eq.~(\ref{cumulantScaling}),
\begin{eqnarray}
\label{expOfCumulantScaling1}
C_R^{(l)} \left( \left( {\rm{\bf x}}, {\rm{\bf x}}^\prime \right) ; \tau , w, \left\{ 
v_k 
\right\} \right) = w^l \left\{ C_l^{(l)} \frac{v_l}{w^l} + C_{l+1}^{(l)} 
\frac{v_{l+1}}{w^{l+1}} + \cdots \right\} \ ,
\end{eqnarray}
with $C_k^{(l)}$ being expansion coefficients depending on ${\rm{\bf x}}$, ${\rm{\bf 
x}}^\prime$, and $\tau$. It is important to recognize that $C_{k<l}^{(l)} = 0$ because 
the corresponding terms are not generated in the perturbation calculation. 
Equation~(\ref{expOfCumulantScaling1}) can be rewritten as
\begin{eqnarray}
\label{expOfCumulantScaling2}
C_R^{(l)} \left( \left( {\rm{\bf x}}, {\rm{\bf x}}^\prime \right) ; \tau , w, \left\{ 
v_k 
\right\} \right) = v_l \left\{ C_l^{(l)} + C_{l+1}^{(l)} \frac{v_{l+1}}{w v_l} + \cdots 
\right\} \ ,
\end{eqnarray}
where the first term on the right hand side gives the leading behavior. Thus $C_R^{(l)}$ 
vanishes upon setting $v_l = 0$ and we cannot gain any further information about 
$C_R^{(l)}$. In particular we cannot determine the associated noise exponent. In other 
words, the $v_{l}$ are dangerously irrelevant in investigating the critical 
properties of the $C_R^{(l\geq 2)}$.

\section{Renormalization Group Analyses}
\label{noisyRGA}

\subsection{Diagrammatic expansion}
\label{noisyDiagrammaticExpansion}
The diagrammatic elements contributing to our renormalization group improved 
perturbation 
calculation are the three point vertex $-g$ and the propagator
\begin{eqnarray}
\label{propagatorDecomp}
\frac{ 1 - \delta_{\stackrel{\mbox{{\tiny $\leftrightarrow$}}}{\lambda}, 
\stackrel{\mbox{{\tiny $\leftrightarrow$}}}{0}}}{{\rm{\bf p}}^2 + \tau + w 
\stackrel{\mbox{{\tiny $\leftrightarrow$}}}{\lambda}^2 + \sum_{l=2}^\infty v_l K_l 
\left( 
\stackrel{\mbox{{\tiny $\leftrightarrow$}}}{\lambda} \right)} 
= \frac{1}{{\rm{\bf p}}^2 + \tau + w \stackrel{\mbox{{\tiny 
$\leftrightarrow$}}}{\lambda}^2 + \sum_{l=2}^\infty v_l K_l \left( 
\stackrel{\mbox{{\tiny 
$\leftrightarrow$}}}{\lambda} \right)} - \frac{\delta_{\stackrel{\mbox{{\tiny 
$\leftrightarrow$}}}{\lambda}, \stackrel{\mbox{{\tiny $\leftrightarrow$}}}{0}}}{{\rm{\bf 
p}}^2 + \tau} \ .
\end{eqnarray}
Note that we have switched to a $\left( {\rm{\bf p}} , \stackrel{\mbox{{\tiny 
$\leftrightarrow$}}}{\lambda} \right)$-representation by employing Fourier 
transformation 
in real and replica space. The notation in Eq.~(\ref{propagatorDecomp}) is somewhat 
symbolic. To treat the irrelevant terms $v_l K_l \left( \stackrel{\mbox{{\tiny 
$\leftrightarrow$}}}{\lambda} \right)$ properly, we have to expand the propagator in a 
power series in the $v_l$ and discard all contributions of higher than linear order in 
the $v_l$. In other words: the irrelevant terms have to be treated as insertions.

Eq.~(\ref{propagatorDecomp}) shows that the principal propagator decomposes into a 
propagator carrying $\stackrel{\mbox{{\tiny $\leftrightarrow$}}}{\lambda}$'s 
(conducting) and one not carrying $\stackrel{\mbox{{\tiny$\leftrightarrow$}}}{\lambda}$'s 
(insulating). This allows for a schematic decomposition of principal diagrams into sums 
of conducting diagrams consisting of conducting and insulating propagators. To two-loop 
order, we obtain the conducting diagrams listed in Fig.~1.

\subsection{Multifractal moments of Feynman diagrams}
\label{multifractalMomentsOfFeynmanDiagrams}
From the decomposition in Sec.~\ref{noisyDiagrammaticExpansion} a real-world 
interpretation of the conducting Feynman diagrams 
emerges\cite{stenull_janssen_oerding_99,janssen_stenull_oerding_99}. They may be viewed 
as resistor networks themselves with conducting propagators corresponding to conductors 
and insulating propagators corresponding to open bonds. The parameters $s$ appearing in 
a Schwinger parametrization of the conducting propagators,
\begin{eqnarray}
\frac{1}{{\rm{\bf p}}^2 + \tau + w \stackrel{\mbox{{\tiny 
$\leftrightarrow$}}}{\lambda}^2 
+ \sum_{l=2}^\infty v_l K_l \left( \stackrel{\mbox{{\tiny $\leftrightarrow$}}}{\lambda} 
\right)} = \int_0^\infty ds \exp \left[ - s \left( {\rm{\bf p}}^2 + \tau + w 
\stackrel{\mbox{{\tiny $\leftrightarrow$}}}{\lambda}^2 + \sum_{l=2}^\infty v_l K_l 
\left( 
\stackrel{\mbox{{\tiny $\leftrightarrow$}}}{\lambda} \right) \right) \right]  \ ,
\end{eqnarray}
correspond to resistances and the replica variables $i\stackrel{\mbox{{\tiny 
$\leftrightarrow$}}}{\lambda}$ to currents. The replica currents are conserved in each 
vertex and we may write for each edge $i$ of a diagram, $\stackrel{\mbox{{\tiny 
$\leftrightarrow$}}}{\lambda}_i = \stackrel{\mbox{{\tiny $\leftrightarrow$}}}{\lambda}_i 
\left( \stackrel{\mbox{{\tiny $\leftrightarrow$}}}{\lambda} , \left\{ 
\stackrel{\mbox{{\tiny $\leftrightarrow$}}}{\kappa} \right\} \right)$, where 
$\stackrel{\mbox{{\tiny $\leftrightarrow$}}}{\lambda}$ is an external current and 
$\left\{ 
\stackrel{\mbox{{\tiny $\leftrightarrow$}}}{\kappa} \right\}$ denotes a complete set of 
independent loop currents. 

The real-world interpretation suggests an effective way of computing the conducting 
diagrams. We learn from the discussion above, that the irrelevant terms have to be 
treated by means of insertions 
\begin{eqnarray}
{\mathcal{O}}^{(l)} = - \frac{1}{2} v_l \int d^d p \, \sum_{\stackrel{\mbox{{\tiny $\leftrightarrow$}}}{\lambda}} K_l \left( \stackrel{\mbox{{\tiny 
$\leftrightarrow$}}}{\lambda} \right) \phi \left( {\rm{\bf p}} , 
\stackrel{\mbox{{\tiny $\leftrightarrow$}}}{\lambda} \right) \phi \left( - {\rm{\bf p}} , - \stackrel{\mbox{{\tiny $\leftrightarrow$}}}{\lambda} \right) \ ,
\end{eqnarray}
where $\phi \left( {\rm{\bf p}} , \stackrel{\mbox{{\tiny $\leftrightarrow$}}}{\lambda} 
\right)$ denotes the Fourier transform of $\varphi \left( {\rm{\bf x}} , 
\stackrel{\mbox{{\tiny $\leftrightarrow$}}}{\theta} \right)$. The resulting diagrams are 
of the type displayed on the left hand side of Fig.~2. We 
express the current dependend part of such a diagram in terms of its power $P$,
\begin{eqnarray}
- s_i v_l \sum_{\big\{ \stackrel{\mbox{{\tiny $\leftrightarrow$}}}{\kappa} \big\}} K_l 
\left( \stackrel{\mbox{{\tiny $\leftrightarrow$}}}{\lambda}_i \right) \exp \bigg[ - w 
\sum_j s_j \stackrel{\mbox{{\tiny $\leftrightarrow$}}}{\lambda}_j^2 \bigg]
=
- s_i v_l \sum_{\big\{ \stackrel{\mbox{{\tiny $\leftrightarrow$}}}{\kappa} \big\}} K_l 
\left( \stackrel{\mbox{{\tiny $\leftrightarrow$}}}{\lambda}_i \right) \exp \left[  w P 
\left( \stackrel{\mbox{{\tiny $\leftrightarrow$}}}{\lambda} , \left\{ 
\stackrel{\mbox{{\tiny $\leftrightarrow$}}}{\kappa} \right\} \right)  \right] \ .
\end{eqnarray}
The summation is carried out by completing the squares in the exponential. The 
corresponding shift in the loop currents is given by the minimum of the quadratic form 
$P$ which is determined by a variation principle completely analogous to the one stated 
in Eq.~(\ref{noisyVariationPrinciple2}). Thus, completing of the squares is equivalent to 
solving Kirchhoff's equations for the diagram. It leads to
\begin{eqnarray}
- s_i v_l \sum_{\big\{ \stackrel{\mbox{{\tiny $\leftrightarrow$}}}{\kappa} \big\}} K_l 
\left( \stackrel{\mbox{{\tiny $\leftrightarrow$}}}{\lambda}_i^{\mbox{\scriptsize ind 
\normalsize}} + \sum_j C_{i,j} \left( \left\{ s \right\} \right) \stackrel{\mbox{{\tiny 
$\leftrightarrow$}}}{\kappa}_j \right) \exp \bigg[ - w R \left( \left\{ s \right\} 
\right) \stackrel{\mbox{{\tiny $\leftrightarrow$}}}{\lambda}^2 - w \sum_{i,j} B_{i,j} 
\left( \left\{ s \right\} \right) \stackrel{\mbox{{\tiny $\leftrightarrow$}}}{\kappa}_i 
\cdot \stackrel{\mbox{{\tiny $\leftrightarrow$}}}{\kappa}_j \bigg]
 \ .
\end{eqnarray}
$\stackrel{\mbox{{\tiny $\leftrightarrow$}}}{\lambda}_i^{\mbox{\scriptsize ind 
\normalsize}} = c_i \left( \left\{ s \right\} \right) \stackrel{\mbox{{\tiny 
$\leftrightarrow$}}}{\lambda}$ is the current induced by the external current into edge 
$i$. $c_i \left( \left\{ s \right\} \right)$ and $ C_{i,j} \left( \left\{ s \right\} 
\right)$ are homogeneous functions of the Schwinger parameters of degree zero. $B_{i,j} 
\left( \left\{ s \right\} \right)$ and the total resistance of the diagram $R \left( 
\left\{ s \right\} \right)$ are homogeneous functions of the Schwinger parameters of 
degree one. By a suitable choice of the $\stackrel{\mbox{{\tiny $\leftrightarrow$}}}{\kappa}_i$ the matrix constituted by the $B_{i,j}$ is rendered diagonal, i.e., $B_{i,j} \sim \delta_{i,j}$. At this stage it is convenient to switch to continuous currents and to replace the summation by an integration,
\begin{eqnarray}
\sum_{\big\{ \stackrel{\mbox{{\tiny $\leftrightarrow$}}}{\kappa} \big\}} \to \int 
\prod_{i=1}^L d \stackrel{\mbox{{\tiny $\leftrightarrow$}}}{\kappa}_i \ , 
\end{eqnarray}
where $d \stackrel{\mbox{{\tiny $\leftrightarrow$}}}{\kappa}$ is an abbreviation for 
$\prod_{\alpha ,\beta =1}^{D,E} d \kappa^{(\alpha ,\beta)}$ and $L$ stands for the 
number of independent conducting loops. This integration is Gaussian and therefore 
straightforward. In the limit $D \to 0$ one obtains
\begin{eqnarray}
\label{huhu}
- s_i v_l K_l \left( \stackrel{\mbox{{\tiny 
$\leftrightarrow$}}}{\lambda}_i^{\mbox{\scriptsize ind \normalsize}} \right) + \cdots
= - s_i c_i \left( \left\{ s \right\} \right)^{2l} v_l K_l \left( \stackrel{\mbox{{\tiny 
$\leftrightarrow$}}}{\lambda} \right) + \cdots \ . 
\end{eqnarray}
The terms neglected in Eq.~(\ref{huhu}) are not required in calculating the $\psi_l$. 
This issue is discussed in detail in Sec.~\ref{noisyRenormalization}. Diagrammatically, 
the calculation scheme can be condensed into Fig.~2.
Appendix~\ref{app:calcSchemeEx} illustrates the calculations in terms of an example.

So far we have inserted ${\mathcal{O}}^{(l)}$ only in one of the conducting 
propagators. However, each of them has to get an insertion. Moreover, the integrations 
over loop momenta and Schwinger parameters remain to be carried out. All in all, each 
diagram can be written as
\begin{eqnarray}
\label{noisyExpansionOfDiagrams}
I \left( {\rm{\bf p}}^2 , \stackrel{\mbox{{\tiny $\leftrightarrow$}}}{\lambda} \right) 
&=& 
I_P \left( {\rm{\bf p}}^2 \right) - I_W \left( {\rm{\bf p}}^2 \right) w 
\stackrel{\mbox{{\tiny $\leftrightarrow$}}}{\lambda}^2 -  I_V^{(l)} 
\left( {\rm{\bf p}}^2 \right) v_l K_l \left( \stackrel{\mbox{{\tiny 
$\leftrightarrow$}}}{\lambda} \right) + \cdots
\nonumber \\
&=& \int_0^\infty \prod_i ds_i \left[ 1 - R \left(  \left\{ s_i \right\} 
\right) w \stackrel{\mbox{{\tiny $\leftrightarrow$}}}{\lambda}^2 -  
C^{(l)} \left(  \left\{ s_i \right\} \right) v_l K_l \left( \stackrel{\mbox{{\tiny 
$\leftrightarrow$}}}{\lambda} \right) + \cdots \right] D \left( {\rm{\bf p}}^2, \left\{ 
s_i \right\} \right) \ .
\end{eqnarray}
Here $D \left( {\rm{\bf p}}^2, \left\{ s_i \right\} \right)$ stands for the integrand 
one 
obtains upon Schwinger parametrization of the corresponding diagram in the usual 
$\phi^3$ 
theory. $C^{(l)} \left(  \left\{ s_i \right\} \right)$ is defined as
\begin{eqnarray}
\label{diagramCumulant}
C^{(l)} \left(  \left\{ s_i \right\} \right) = \sum_i s_i \, c_i \left( \left\{ s 
\right\} \right)^{2l} = \sum_i s_i \left( \stackrel{\mbox{{\tiny 
$\leftrightarrow$}}}{\lambda}_i^{\mbox{\scriptsize ind \normalsize}} \hspace{-2mm}/ 
\stackrel{\mbox{{\tiny $\leftrightarrow$}}}{\lambda}  \right)^{2l}  \ ,
\end{eqnarray}
where the sum runs over all conducting propagators of the diagram. Notice the analogy of 
the $C^{(l)} \left(  \left\{ s_i \right\} \right)$ to the generalized multifractal 
moments we introduce in App.~\ref{app:Generalization}. Thus, we refer to the $C^{(l)} 
\left(  \left\{ s_i \right\} \right)$ as multifractal moments of conducting Feynman 
diagrams.

\subsection{Renormalization and scaling}
\label{noisyRenormalization}
By employing dimensional regularization and minimal subtraction we proceed with standard 
techniques of renormalized field theory\cite{amit_zinn-justin}. The renormalization of 
the $v_l$, however, involves some peculiarities that we will discuss in this section.

An operator ${\mathcal{O}}_i$ of a given naive dimension $\left[ {\mathcal{O}}_i \right]$ inserted one time in a vertex function generates in general new 
primitive divergencies corresponding to all operators of equal or lower naive dimension. 
Thus, one needs these newly generated operators as counterterms in the Hamiltonian.

The operators of lower naive dimension can be isolated by additive renormalization,
\begin{eqnarray}
{\mathcal{O}}_i \to \hat{{\mathcal{O}}}_i = {\mathcal{O}}_i 
- \sum_{\left[ {\mathcal{O}}_j \right] < \left[ {\mathcal{O}}_i 
\right]} X_{i,j} {\mathcal{O}}_j \ .
\end{eqnarray}
Dimensional regularization in conjunction with minimal subtraction leads to $X_{i,j}$ 
containing at least a factor $\tau$. These $X_{i,j}$ vanish at the critical point. Hence, 
the operators of lower naive dimension will not be considered in the following.

As argued in Sec.~\ref{multifractalMomentsOfFeynmanDiagrams} the term proportional to 
$v_l$ in Eq.~(\ref{noisyExpansionOfDiagrams}) is generated by inserting the operator 
${\mathcal{O}}^{(l)}$. Inserting ${\mathcal{O}}^{(l)}$ into a 
diagram with $n$ external legs, see Fig.~3,
generates primitive divergencies which must be cancelled by counter terms of the  
structure
\begin{eqnarray}
\label{generalOperator}
P_r \left( \stackrel{\mbox{{\tiny $\leftrightarrow$}}}{\lambda} \right) {\rm{\bf 
p}}^{2a} \phi \left( {\rm{\bf p}} , \stackrel{\mbox{{\tiny $\leftrightarrow$}}}{\lambda} 
\right)^n \ ,
\end{eqnarray}
where
\begin{eqnarray}
P_r \left( \stackrel{\mbox{{\tiny $\leftrightarrow$}}}{\lambda} \right) = \prod_i K_i 
\left( \stackrel{\mbox{{\tiny $\leftrightarrow$}}}{\lambda} \right)^{r_i} \ ,
\end{eqnarray}
with $\sum_i i r_i = r$, is a homogeneous polynomial of degree $2r$. Note, that the 
notation we use here and in the following is symbolic, since such a counter term has to 
depend on the entire set of external momenta and currents. $\phi \left( {\rm{\bf p}} , 
\stackrel{\mbox{{\tiny $\leftrightarrow$}}}{\lambda} \right)^n$ is for example an 
abbreviation for $\prod_{i=1}^n \phi \left( {\rm{\bf p}}_i , \stackrel{\mbox{{\tiny 
$\leftrightarrow$}}}{\lambda}_i \right)$.

The leading contribution comes from operators having the same naive dimension as 
${\mathcal{O}}^{(l)}$, i.e., those satisfying
\begin{eqnarray}
\label{blabla}
2 \left( l+2-3 \right) = 2 \left( r+a+n-3 \right) \ .
\end{eqnarray}
Here we expressed the naive dimension with help of Eq.~(\ref{naiveDimOfInsertion}). For $n=2$ one is led to $l \geq r$, i.e., the insertion of ${\mathcal{O}}^{(l)}$ generates operators containing homogeneous polynomials in the replica currents of degree equal or lower $2l$. In particular ${\mathcal{O}}^{(l)}$ generates an operator of type
\begin{eqnarray}
v_l K_l \left( \stackrel{\mbox{{\tiny $\leftrightarrow$}}}{\lambda} \right) \phi \left( 
{\rm{\bf p}} , \stackrel{\mbox{{\tiny $\leftrightarrow$}}}{\lambda} \right)^2 \ .
\end{eqnarray}
The important question now is, if the other operators generated by ${\mathcal{O}}^{(l)}$ generate operators of this type, too. Consider $n \geq 3$. With help 
of Eq.~(\ref{blabla}) one obtains $l-1 \geq r \geq 1$, where the second inequality is a 
consequence of the limit $D\to 0$. Bearing in mind that maximal homogeneous polynomials 
of degree $l-1$ in $\stackrel{\mbox{{\tiny $\leftrightarrow$}}}{\lambda}$ are generated, 
we reinsert these operators of the type in Eq.~(\ref{generalOperator}) with $n \geq 3$ 
into two-leg diagrams, see Fig.~4. The resulting terms are of the form
\begin{eqnarray}
P_{r^\prime} \left( \stackrel{\mbox{{\tiny $\leftrightarrow$}}}{\lambda} \right) 
{\rm{\bf p}}^{2a^\prime} \phi \left( {\rm{\bf p}} , \stackrel{\mbox{{\tiny 
$\leftrightarrow$}}}{\lambda} \right)^2 \ ,
\end{eqnarray}
with the leading contributions satisfying $r+a+n-3=r^\prime+a^\prime-1$. Thus, $r^\prime 
\geq r + a - a^\prime + 1$, i.e., the homogeneous polynomials in $\stackrel{\mbox{{\tiny 
$\leftrightarrow$}}}{\lambda}$ may have a higher degree than $2l$. However, they are of 
the type
\begin{eqnarray}
P_{r^\prime} \left( \stackrel{\mbox{{\tiny $\leftrightarrow$}}}{\lambda} \right) = 
K_{1} \left( \stackrel{\mbox{{\tiny $\leftrightarrow$}}}{\lambda} \right)^s 
\prod_{2 \leq i \leq r} K_i \left( \stackrel{\mbox{{\tiny $\leftrightarrow$}}}{\lambda} 
\right)^{r_i} \ ,
\end{eqnarray}
with $\sum_i i r_i \leq r \leq l-1$ and $\sum_i i r_i + s = r^\prime$. These 
polynomials have a higher symmetry than the original $K_l$. 

We conclude that ${\mathcal{O}}^{(l)}$ generates itself and an entire family 
of new operators but these in turn do not generate ${\mathcal{O}}^{(l)}$. In 
principle, the entire family of operators associated with ${\mathcal{O}}^{(l)}$ has to be taken into account in the renormalization proceedure, 
leading to a renormalization in matrix form
\begin{eqnarray}
\underline{\hat{{\mathcal{O}}}}^{(l)} \to \underline{{\mathaccent"7017 {\mathcal{O}}}}^{(l)} = \underline{\underline{Z}}^{(l)} \underline{\hat{{\mathcal{O}}}}^{(l)} \ .
\end{eqnarray}
The vector
\begin{eqnarray}
\underline{\hat{{\mathcal{O}}}}^{(l)} = \left( {\mathcal{O}}^{(l)}, 
\hat{{\mathcal{O}}}_2^{(l)}, \cdots \right)
\end{eqnarray}
contains the family associated with ${\mathcal{O}}^{(l)}$. For the remaining 
renormalizations, we employ the same scheme as in~\cite{stenull_janssen_oerding_99},
\begin{mathletters}
\label{renormalizationScheme}
\begin{eqnarray}
\varphi \to {\mathaccent"7017 \psi} = Z^{1/2} \varphi \ ,& & \
\tau \to {\mathaccent"7017 \tau} = Z^{-1} Z_{\tau} \tau \ ,
 \\
w \to {\mathaccent"7017 w} = Z^{-1} Z_{w} w \ , & & \
g \to {\mathaccent"7017 g} = Z^{-3/2} Z_u^{1/2} G_\epsilon^{-1/2} u^{1/2} 
\mu^{\epsilon /2} \ .
\end{eqnarray}
\end{mathletters}
In Eq.~(\ref{renormalizationScheme}) $\epsilon$ stands for $6-d$ and the factor 
$G_\epsilon = (4\pi )^{-d/2}\Gamma (1 + \epsilon /2)$, with $\Gamma$ denoting the Gamma 
function, is introduced for convenience.

According to the arguments given above the renormalization matrix
\begin{eqnarray}
\underline{\underline{Z}}^{(l)} = \underline{\underline{1}} + {\sl O} \left( u \right) 
\end{eqnarray}
has a particularly simple 
structure,
\begin{eqnarray}
\underline{\underline{Z}}^{(l)} = 
\left(
\begin{array}{cccc}
Z^{(l)} & \Diamond & \cdots & \Diamond \\
0       & \Diamond & \cdots & \Diamond \\
\vdots  & \vdots & \ddots & \vdots \\
0       & \Diamond & \cdots & \Diamond
\end{array}
\right) \ ,
\end{eqnarray}
$\underline{\underline{1}}$ stands for the unit matrix and $\Diamond$ symbolizes elements that we do not evaluate. In this paper, we determine $Z^{(l)}$ to the order of two loops. $Z$, $Z_\tau$ and $Z_u$ are the usual Potts model $Z$ factors. They have been computed to three-loop order by de Alcantara Bonfim {\it et al}\cite{alcantara_80}. $Z_w$ is known to two-loop order\cite{stenull_janssen_oerding_99}. 

The unrenormalized theory has to be independent of the length scale $\mu^{-1}$ 
introduced by renormalization. In particular, the connected $N$ point 
correlation functions with an insertion of $\underline{\hat{{\mathcal{O}}}}^{(l)}$ must be independent of $\mu$, i.e.,  
\begin{eqnarray}
\label{noisyIndependence}
\mu \frac{\partial}{\partial \mu} {\mathaccent"7017 G}_N \left( \left\{ {\rm{\bf 
x}}, {\mathaccent"7017 w} \stackrel{\mbox{{\tiny $\leftrightarrow$}}}{\lambda}^2, 
 \right\} ; {\mathaccent"7017 \tau}, {\mathaccent"7017 g} \right)_{\underline{\hat{{\mathcal{O}}}}^{(l)}}  = 0
\end{eqnarray}
for all $N$. Eq.~(\ref{noisyIndependence}) translates via the Wilson functions
\begin{mathletters}
\begin{eqnarray}
\label{noisyWilson}
\beta \left( u \right) = \mu \frac{\partial u}{\partial \mu} \bigg|_0 \ ,
& & \
\kappa \left( u \right) = \mu \frac{\partial
\ln \tau}{\partial \mu}  \bigg|_0 \ ,
 \\
\zeta \left( u \right) = \mu \frac{\partial \ln w}{\partial \mu}  \bigg|_0 \ ,
& & \
\gamma \left( u \right) = \mu \frac{\partial }{\partial \mu} \ln Z  
\bigg|_0 \ ,
\\
\underline{\underline{\gamma}}^{(l)} \left( u \right) &=& - \mu \frac{\partial 
}{\partial \mu} \ln \underline{\underline{Z}}^{(l)}  \bigg|_0 \ ,
\end{eqnarray}
\end{mathletters}
where the bare quantities are kept fix while taking the derivatives, into the 
Gell-Mann-Low renormalization group equation
\begin{eqnarray}
\label{noisyRGG}
\lefteqn{ \left\{ \left[ \mu \frac{\partial }{\partial \mu} + \beta \frac{\partial 
}{\partial u} + \tau \kappa \frac{\partial }{\partial \tau} + w \zeta \frac{\partial 
}{\partial w} + \frac{N}{2} \gamma 
\right] \underline{\underline{1}} + \underline{\underline{\gamma}}^{(l)} \right\} }
\nonumber \\
&& \times \, 
G_N \left( \left\{ {\rm{\bf x}} ,w \stackrel{\mbox{{\tiny 
$\leftrightarrow$}}}{\lambda}^2 \right\} ; \tau, u, \mu \right)_{\underline{\hat{\mathcal{O}}}^{(l)}} = 0 \ .
\end{eqnarray}
The particular form of the Wilson functions can be extracted from the renormalization scheme and the $Z$ factors. At the infrared stable fixed point $u^\ast$, determined by $\beta \left( u^\ast \right) = 0$, the renormalization group equation reduces to
\begin{eqnarray}
\label{fixNoisyRGG}
\lefteqn{ \left\{ \left[ \mu \frac{\partial }{\partial \mu} + \tau \kappa^\ast \frac{\partial }{\partial \tau} + w \zeta^\ast \frac{\partial 
}{\partial w} + \frac{N}{2} \gamma^\ast \right] \underline{\underline{1}} + \underline{\underline{\gamma}}^{(l)\ast} \right\} }
\nonumber \\
&& \times \, 
G_N \left( \left\{ {\rm{\bf x}} ,w \stackrel{\mbox{{\tiny 
$\leftrightarrow$}}}{\lambda}^2 \right\} ; \tau, u^\ast, \mu \right)_{\underline{\hat{\mathcal{O}}}^{(l)}} = 0 \ ,
\end{eqnarray}
where $\gamma^\ast = \gamma \left( u^\ast \right)$, $\kappa^\ast = \kappa \left( u^\ast 
\right)$, $\zeta^\ast = \zeta \left( u^\ast \right)$, and $\underline{\underline{\gamma}}^{(l)\ast} = \underline{\underline{\gamma}}^{(l)} \left( u^\ast \right)$.

The matrix $\underline{\underline{\gamma}}^{(l)}$ inherits the simple structure of 
$\underline{\underline{Z}}^{(l)}$, 
\begin{eqnarray}
\underline{\underline{\gamma}}^{(l)} = 
\left(
\begin{array}{cccc}
\gamma^{(l)} & \Diamond & \cdots & \Diamond \\
0       & \Diamond & \cdots & \Diamond \\
\vdots  & \vdots & \ddots & \vdots \\
0       & \Diamond & \cdots & \Diamond
\end{array}
\right) \ .
\end{eqnarray}
Owing to this structure, $\left| 1 \right\rangle = \left( 1, 0, \cdots , 0 \right)^T$ is 
a right eigenvector of $\underline{\underline{\gamma}}^{(l)\ast}$ with eigenvalue 
$\gamma^{(l)\ast}$. The remaining right eigenvectors with eigenvalues $\gamma_k^{(l)\ast}$, $k 
\geq 2$, we denote by $\left| k \right\rangle$. The left eigenvectors of 
$\underline{\underline{\gamma}}^{(l)\ast}$ are $\left\langle 1 \right| = \left( 1, \Diamond, 
\cdots , \Diamond \right)$ and $\left\langle k \right| = \left( 0, \Diamond, \cdots , \Diamond 
\right)$. In terms of the eigenvectors, $\underline{\underline{\gamma}}^{(l)\ast}$ can be 
spectrally decomposed into
\begin{eqnarray}
\label{spectralDeco}
\underline{\underline{\gamma}}^{(l)\ast} = \left| 1 \right\rangle \gamma^{(l)\ast} \left\langle 1 \right| + \sum_{k \geq 2} \left| k \right\rangle \gamma^{(l)\ast}_k \left\langle k \right|
 \ .
\end{eqnarray}

Now it is important to realize, that
\begin{eqnarray}
\left\langle 1 \right| \underline{\hat{\mathcal{O}}}^{(l)} &=& \hat{\mathcal{O}}^{(l)} + \sum_{k \geq 2} \Diamond \, \hat{\mathcal{O}}^{(l)}_k = {\mathcal{A}}^{(l)} \ ,
\nonumber \\
\left\langle k \right| \underline{\hat{\mathcal{O}}}^{(l)} &=&  \sum_{k \geq 
2} \Diamond \, \hat{\mathcal{O}}^{(l)}_k \ ,
\end{eqnarray}
i.e., only ${\mathcal{A}}^{(l)}$ contains the operator ${\mathcal{O}}^{(l)}$ we are interested in. We substitute Eq.~(\ref{spectralDeco}) into the renormalization group equation (\ref{fixNoisyRGG}) and act on the entire equation with $\left\langle 1 \right|$. The 
result is 
\begin{eqnarray}
\label{noisyRGGsimpler}
\left[ \mu \frac{\partial }{\partial \mu} + \tau \kappa^\ast \frac{\partial }{\partial \tau} + w \zeta^\ast \frac{\partial 
}{\partial w} + \frac{N}{2} \gamma + \gamma^{(l)}\right] 
G_N \left( \left\{ {\rm{\bf x}} ,w \stackrel{\mbox{{\tiny 
$\leftrightarrow$}}}{\lambda}^2 \right\} ; \tau, u^\ast, \mu \right)_{{\mathcal{A}}^{(l)}} = 0 \ .
\end{eqnarray}
Equation~(\ref{noisyRGGsimpler}) is solved by the method of characteristics. The solution reads
\begin{eqnarray}
\label{noisySolOfRgg}
G_N \left( \left\{ {\rm{\bf x}} ,w \stackrel{\mbox{{\tiny 
$\leftrightarrow$}}}{\lambda}^2 \right\} ; \tau, u^\ast, \mu \right)_{{\mathcal{A}}^{(l)}} = 
\varrho^{\gamma^\ast N/2 + \gamma^{(l)\ast}} G_N \left( \left\{ \varrho{\rm{\bf x}} ,\varrho^{\zeta^\ast}w 
\stackrel{\mbox{{\tiny $\leftrightarrow$}}}{\lambda}^2 \right\} ; 
\varrho^{\kappa^\ast}\tau , u^\ast, \varrho \mu \right)_{{\mathcal{A}}^{(l)}} \ . 
\end{eqnarray}

To derive a scaling relation for the correlation functions, a dimensional analysis 
remains to be performed. It yields
\begin{eqnarray}
\label{noisyDimAna}
\lefteqn{ G_N \left( \left\{ {\rm{\bf x}} ,w \stackrel{\mbox{{\tiny 
$\leftrightarrow$}}}{\lambda}^2 \right\} ; \tau, u, \mu \right)_{{\mathcal{A}}^{(l)}} }
\nonumber \\
& & =
\mu^{(d-2) N/2 -2} G_N \left( \left\{ \mu {\rm{\bf x}} ,\mu^{-2} w \stackrel{\mbox{{\tiny 
$\leftrightarrow$}}}{\lambda}^2 \right\} ; \mu^{-2}\tau , u, 1 \right)_{{\mathcal{A}}^{(l)}} \  . 
\end{eqnarray}
From Eqs.~(\ref{noisySolOfRgg}) and (\ref{noisyDimAna}) we deduce the scaling behavior
\begin{eqnarray}
\label{noisyScalingRel}
\lefteqn{ G_N \left( \left\{ {\rm{\bf x}} ,w \stackrel{\mbox{{\tiny 
$\leftrightarrow$}}}{\lambda}^2 \right\} ; \tau, u, \mu \right)_{{\mathcal{A}}^{(l)}} }
\nonumber \\
& & = 
\varrho^{(d-2+\eta)N/2 -\psi_l/\nu} G_N \left( \left\{ \varrho{\rm{\bf x}} , \varrho^{-\phi/\nu} w 
\stackrel{\mbox{{\tiny $\leftrightarrow$}}}{\lambda}^2 \right\} ; 
\varrho^{-1/\nu}\tau , u^\ast, \mu \right)_{{\mathcal{A}}^{(l)}} \ .
\end{eqnarray}
$\eta = \gamma^\ast$ and $\nu = \left( 2 - \kappa^\ast \right)^{-1}$ are the well known 
critical exponents for percolation. They are known to third order in 
$\epsilon$\cite{alcantara_80}:
\begin{eqnarray}
\label{exponentEta}
\eta = \gamma^\ast = - \frac{1}{21}\epsilon - 
\frac{206}{9261}\epsilon^2 + \left[ - \frac{93619}{8168202} + \frac{256}{7203} 
\zeta \left( 3 \right) \right]\epsilon^3 + {\sl O} \left( \epsilon^4  \right) \ ,
\end{eqnarray}
and
\begin{eqnarray}
\label{exponentNu}
\nu = \left( 2 - \kappa^\ast \right)^{-1} = \frac{1}{2} + 
\frac{5}{84}\epsilon + \frac{589}{37044}\epsilon^2 + \left[ 
\frac{716519}{130691232} - \frac{89}{7203} \zeta \left( 3 \right) 
\right]\epsilon^3 + {\sl O} \left( \epsilon^4  \right) \ .
\end{eqnarray}
Note that $\zeta$ in Eqs.~(\ref{exponentEta}) and (\ref{exponentNu}) stands for the 
Riemann zeta function and should not be confused with the Wilson function defined above. 
$\phi = \nu \left( 2 - \zeta^\ast \right)$ is the resistance exponent known to second 
order in $\epsilon$\cite{lubensky_wang_85,stenull_janssen_oerding_99},
\begin{eqnarray}
\label{exponentPhi}
\phi = \nu \left( 2 - \zeta^\ast \right) = 1 + \frac{1}{42}\epsilon + 
\frac{4}{3087}\epsilon^2 + {\em O} \left( \epsilon^3  \right) \ . 
\end{eqnarray}
The noise exponents $\psi_l$ are defined by $\psi_l = \nu \left( 2 - \gamma^{(l)\ast} 
\right)$. The expansion of $\psi_l$ to second order in $\epsilon$ is given below.

Now we are in the position to derive the scaling behavior of $C_R^{(l)}$. From Eq.~(\ref{noisyScalingRel}) we find upon choosing $\varrho = |{\bf x}-{\bf x}^\prime |^{-1}$ and Taylor expanding that the two point correlation function $G=G_2$ scales at criticality
as
\begin{eqnarray}
\lefteqn{ G \left( {\bf x}, {\bf x}^\prime ; \stackrel{\mbox{{\tiny 
$\leftrightarrow$}}}{\lambda} \right) = |{\bf x}-{\bf x}^\prime 
|^{2-d-\eta} }
\nonumber \\
& & \times \left\{ 1 + w \stackrel{\mbox{{\tiny $\leftrightarrow$}}}{\lambda}^2 |{\bf 
x}-{\bf x}^\prime |^{\phi/\nu} + v_l K_l \left( \stackrel{\mbox{{\tiny 
$\leftrightarrow$}}}{\lambda} \right) |{\bf x}-{\bf x}^\prime |^{\psi_l/\nu} + \cdots 
\right\} \ ,
\end{eqnarray}
where we have dropped several arguments for notational simplicity. With 
Eq.~(\ref{exploitGenFkt}) the desired scaling behavior of $C_R^{(l)}$ is now readily 
obtained as 
\begin{eqnarray}
\label{kommi2}
C_R^{(l)} \sim |{\bf x}-{\bf x}^\prime |^{\psi_l/\nu} \ .
\end{eqnarray}

At this point, we emphasize once more the outstanding role of the ${\mathcal{O}}^{(l)}$, which warrants calling them master operators\cite{stenull_janssen_epl2000}. Each multifractal moment $M^{(l)}_I$ has a master operator as field theoretic counterpart. The master operators are highly and dangerously irrelevant in the renormalization group sense. Therefore, each master operator needs in general a myriad of other irrelevant operators for renormalization. However, the renormalization of these servant operators does not induce their master. It follows, that the servant operators can be neglected in determining the scaling index of their master operator, i.e., one is spared the computation and diagonalization of giant renormalization matrices.

Our $\epsilon$-expansion result for the noise exponents reads
\begin{eqnarray}
\label{monsterExponent}
\psi_l &=& 1 + \frac{\epsilon}{7  \left( 1+l \right) \left( 1+2l \right)} + 
\frac{\epsilon^2}{12348
      \left( 1 + l \right)^3
      \left( 1 + 2l \right)^3}
\nonumber \\
&\times& \Bigg\{
 313 - 672\gamma + 
        l\bigg\{ 3327 - 4032\gamma - 
           8l\Big\{ 4
               \left( -389 + 273\gamma
                 \right)    
\nonumber \\
&+&   
              l\left[ -2076 + 1008\gamma + 
               l  \left( -881 + 336\gamma
                     \right) \right]  \Big\}  \bigg\}
\nonumber \\                      
            &-& 
        672\left( 1 + l \right)^2
         \left( 1 + 2l \right)^2
         \Psi(1 + 2l)  \Bigg\} + {\em O} \left( \epsilon^3  \right)
\end{eqnarray}
in agreement to first order in $\epsilon$ with the one-loop calculation by PHL. $\gamma = 0.577215...$ denotes Euler's constant and $\Psi$ stands for the Digamma function\cite{abramowitz_stegun_65}. Equation~(\ref{monsterExponent}) is valid not only for $l \geq 2$ since it can be continued analytically down to $l=0$. A plot of $\psi_l$ versus $\epsilon$ is given in Fig.~5. We point out that Eq.~(\ref{monsterExponent}) evaluated at $l=1$ is in conformity with the result for $\phi$ stated in Eq.~(\ref{exponentPhi}), i.e., our result for $\psi_l$ satisfies an important consistency check steaming from $C^{(1)}_R = M^{(1)}_R$. Blumenfeld {\em et al.}\cite{blumenfeld_etal_87} proved that $\psi_l$ is a convex monotonically decreasing function of $l$. Note from Fig.~5 that our result for $\psi_l$ captures this feature for reasonable values of $\epsilon$. It reduces to unity in the limit $l\to \infty$ as one expects from the relation of $\psi_\infty$ to the fractal dimension of the red bonds (see App.~\ref{app:relationToTheBackboneDimension}). Moreover, analytic continuation of $\psi_l$ to $l=0$ shows that $\psi_{0} = \nu D_B$ up to order ${\em O} \left( \epsilon^3  \right)$ as expected (see App.~\ref{app:relationToTheBackboneDimension}).

\section{Comparison to numerical data}
In this section we compare our result for $\psi_l$ to numerical values. Instead of working with $\psi_l$ directly we compare $\psi_l /\nu$ because data for exponents of this type is available in the literature. For the comparison it is not sufficient to simply evaluate Eq.~(\ref{monsterExponent}) at $\epsilon =3$ or $\epsilon =4$. The pure $\epsilon$-expansion gives for small spatial dimension poor quantitative predictions. However, one can improve the $\epsilon$-expansion by incorporating rigorously known features. We carry out a rational approximation which takes into account that $\psi_l /\nu = 1$ in one dimension. Practically this is done by adding an appropriate third order term to the $\epsilon$-expansion of $\psi_l /\nu$. We refrain from stating the so obtained formula explictly because it is a little lengthy. Instead we plot it for $\epsilon = 3$, i.e. $d=3$, in Fig.~6. Our analytic result shows remarkable agreement with the the available numerical data for $d=3$\cite{moukarzel_98,batrouni&co_96}. For $l=0$ our result lays slightly outside the error bars of the simulations. However, the deviation of the values is less than $3\%$. For $l=1,2,3$ our result is within the error bars of the simulation. There are also numerical values available for $d=2$\cite{grassberger_99,rammal_etal_85}. Here, however, the agreement is much less pronounced. For $l=0,1,2,3,4$ we find a deviation of the order of $30\%$. It appears that the dependence of $\psi_l$ on dimensionality is too rich in structure to be approximated well at $d=2$ by a series of a few terms.

\section{Conclusions}
\label{noisyConclusions}
We studied the multifractal moments of the current distribution in RRN by renormalized field theory. Our approach thrived on two cornerstones. First, the Feyman digrams for RRN can be interpreted as being resistor networks themselves. In this paper we extended our real-world interpretation by introducing multifractal moments for Feynman diagrams. The real-world interpretation proves to be a powerful tool which allows a highly efficient calculation of the diagrams. 

The second cornerstone was our concept of master operators. Whereas the field theoretic operator associated with the resistance exponent $\phi$ is relevant in the renormalization group sense, the operators associated with the $\psi_{l\geq 2}$ are dangerously irrelevant master operators. Due to their irrelevance the master operators generate a multitude of other irrelevant operators, the servants, which in principle must all be taken into account in the renormalization procedure. The servants however, do not influence the scaling index of their master. Without this property, one would have to compute and diagonalize entire renormalization matrices for determining the $\psi_{l}$. These renormaliztion matrizes are giants for large $l$. Without the master property it would be practially impossible to compute the $\psi_{l}$ for arbitrary $l$.

To our knowledge this is the first time that an entire family of multifractal exponents has been calculated to two-loop order, at least for percolation. Our result is for dimensions near the upper critical dimension 6 the most accurate analytic estimate for the $\psi_{l}$ that we know of. It fulfils several consistency checks. Moreover, it agrees remarkably well with numerical data for $d=3$. As one expects, the agreement suffers by further decreasing the dimension. The dependence of the $\psi_l$ on dimensionality appears to be too complex for being approximated well at $d=2$ by a series of a few terms.

We expect that our concept of master operators can be applied to other systems showing multifractality. It works for example to describe the moments of the current distribution in random resistor diode networks. A two-loop calculation of the corresponding family of mutifractal exponents will be reported in the near future\cite{stenull_janssen_dirMult_2000}. Another example for the applicability of the concept of master operators is the problem of diffusion near polymers. For this problem von Ferber and Holovatch\cite{vonFerber_holovatch} formulated a field theory which comprises dangerously irrelevant operators. Due to the symmetry properties of their operators no other irrelevant operators are generated in the perturbation calculations of these authors. Thus, the operators studied by von Ferber and Holovatch are particular simple instances of master operators. Their scaling index is not influenced by any other operator simply because the number of their servants is zero. In this sense these master operators may be called poor.

It might turn out
that the field theoretic operators associated with multifractal quantities are in general master operators. In this case the concept of master operators would be a key in understanding the origin of multifractality, at least from a field theoretic point of view.

\acknowledgements
We acknowledge support by the Sonderforschungsbereich 237 ``Unordnung und 
gro{\ss}e Fluktuationen'' of the Deutsche Forschungsgemeinschaft. 

\appendix
\section{Generalization of Cohn's Theorem and Generalized Multifractal Moments}
\label{app:Generalization}
Consider a generalized power of the form
\begin{eqnarray}
\label{generalizedPower}
P \left( \left\{ I_b \right\} \right) = \sum_b \rho_b F \left( I_b \right) \ ,
\end{eqnarray}
where $F$ is some function of bond currents $I_b$. As argued in Sec.~\ref{noisyModel}, $I_b$ is in 
general a function of the external current $I$ and a complete set of loop currents 
$\left\{ I^{(l)} \right\}$. We can exploit the variation 
principle~(\ref{noisyVariationPrinciple2}) to eliminate the loop currents. As a result 
we obtain the $I_b$ as a function of $I$ and $\left\{ \rho_b \right\}$ only. The 
solutions we denote by $I_b^{\mbox{{\scriptsize ind}}}$. The derivative of the so 
obtained power $P \left( \left\{ I_b^{\mbox{{\scriptsize ind}}} \right\} \right)$ with 
respect to a bond resistance $\rho_b$ reads
\begin{eqnarray}
\frac{ \partial P \left( \left\{ I_b^{\mbox{{\scriptsize ind}}} \right\} 
\right)}{\partial \rho_b} = F 
\left( I_b^{\mbox{{\scriptsize ind}}} \right) + \sum_l \sum_{b^\prime} \rho_{b^\prime} 
\frac{ \partial F 
\left( I_{b^\prime}^{\mbox{{\scriptsize ind}}} \right) }{\partial I^{(l)}} \frac{\partial 
I^{(l)} 
}{\partial \rho_b} \ .
\end{eqnarray}
The second term on the right hand side vanishes by virtue of the variation 
principle~(\ref{noisyVariationPrinciple2}),  
\begin{eqnarray}
\frac{ \partial P \left( \left\{ I_b^{\mbox{{\scriptsize ind}}} \right\} 
\right)}{\partial I^{(l)}} = 
\sum_b \rho_b \frac{ \partial F \left( I_{b^\prime}^{\mbox{{\scriptsize ind}}} 
\right)}{\partial 
I^{(l)}} 
= 0 \ .
\end{eqnarray}
Renaming $I_b = I_b^{\mbox{{\scriptsize ind}}}$ we finally obtain
\begin{eqnarray}
\frac{ \partial P \left( \left\{ I_b \right\} \right)}{\partial \rho_b} = F \left( I_b 
\right) 
\end{eqnarray}
as a generalization of Cohn's theorem. For $F \left( I_b \right) = I_b^2$ one retrieves 
the original theorem
\begin{eqnarray}
\label{originalCohn}
\frac{\partial R (x ,x^\prime )}{\partial \rho_{b}} = \left( \frac{I_b}{I} \right)^2 \ .
\end{eqnarray}

Having generalized the power it is natural to generalize the multifractal moments as 
well. Consider the cumulants of the generalized power,  
\begin{eqnarray}
\left\{ P^n \right\}_f^{(c)} = \frac{\partial^n}{\partial \lambda^n } \ln \left\{ \exp 
\left( \lambda P \right) \right\}_f \ .
\end{eqnarray}
In analogy to the resistance cumulants one finds for the leading behavior in the limit 
$s 
\to 0$, 
\begin{eqnarray}
\left\{ P^n \right\}_f^{(c)} &=&  \sum_b \left( \left. \frac{\partial P}{\partial 
\rho_b} 
\right|_{\overline{\rho}} \right)^n \left\{ \delta \rho_b^n  \right\}_f^{(c)} 
\nonumber \\
&=& \sum_b F \left( I_b \right)^n \left\{ \delta \rho_b^n  \right\}_f^{(c)} \ .
\end{eqnarray}
For networks like those described in Sec.~\ref{noisyModel} one has $\left\{ \delta 
\rho_b^n  \right\}_f^{(c)} =v_n$. However, in a more general situation the individual 
bonds may be composed of a series of elementary resistors. The elementary resistors are assumed 
to have independently and identically distributed resistances with mean 
$\overline{\rho}$ 
and higher cumulants $v_{l\geq 2}$. Then
\begin{eqnarray}
\left\{ \delta \rho_b^n  \right\}_f^{(c)}  = n_b v_n = 
\frac{\overline{\rho}_b}{\overline{\rho}} v_n \ ,
\end{eqnarray}
where $n_b$ denotes the number of elementary resistors constituting bond $b$ and 
$\overline{\rho}_b$ is the average resistance of that bond. Upon incorporating a factor 
$\overline{\rho}^{-1}$ into the constants $v_n$ we finally find
\begin{eqnarray}
\left\{ P^n \right\}_f^{(c)} = v_n \sum_b \overline{\rho}_b F \left( I_b \right)^n  \ ,
\end{eqnarray}
with $\sum_b \overline{\rho}_b F \left( I_b \right)^n$ being the $n$th multifractal 
moment of $F \left( I_b \right)$.

\section{Computation of Diagrams I}
\label{app:calcSchemeEx}
In this appendix we illustrate the calculation scheme sketched in 
Sec.~\ref{multifractalMomentsOfFeynmanDiagrams} at the instance of diagram A. For the 
sake of simplicity, we focus on its contribution to the renormalization of $v_2$. We 
neglect all other parts and obtain 
\begin{eqnarray}
\mbox{A} &=& - \int_0^\infty ds_1 ds_2 D \left( {\rm{\bf p}}^2, \left\{ s_i \right\} 
\right) \sum_{\stackrel{\mbox{{\tiny $\leftrightarrow$}}}{\kappa}} \exp \left[ w P 
\left( 
\stackrel{\mbox{{\tiny $\leftrightarrow$}}}{\lambda} , \stackrel{\mbox{{\tiny 
$\leftrightarrow$}}}{\kappa} \right) \right]
\nonumber \\
&\times& \left\{ s_1 v_2 K_2 \left( \stackrel{\mbox{{\tiny 
$\leftrightarrow$}}}{\kappa} \right) + s_2 v_2 K_2 \left( 
\stackrel{\mbox{{\tiny $\leftrightarrow$}}}{\kappa} + \stackrel{\mbox{{\tiny 
$\leftrightarrow$}}}{\lambda} \right) \right\} \ ,
\end{eqnarray}
where $P \left( \stackrel{\mbox{{\tiny $\leftrightarrow$}}}{\lambda} ,  
\stackrel{\mbox{{\tiny $\leftrightarrow$}}}{\kappa} \right) = - s_1 
\stackrel{\mbox{{\tiny $\leftrightarrow$}}}{\kappa}^2 - s_2 \left( 
\stackrel{\mbox{{\tiny 
$\leftrightarrow$}}}{\kappa} + \stackrel{\mbox{{\tiny $\leftrightarrow$}}}{\lambda} 
\right)^2$ is, according to the real-world interpretation, the power of the diagram and 
$D 
\left( {\rm{\bf p}}^2, \left\{ s_i \right\} \right)$ stands for
\begin{eqnarray} 
D \left( {\rm{\bf p}}^2, \left\{ s_i \right\} \right) = \frac{g^2}{2} \int_{\rm{\bf q}} 
\exp \left[ - 
\left( s_1 + s_2 \right) \tau - s_1 {\rm{\bf q}}^2 - s_2 \left( {\rm{\bf q}} + {\rm{\bf 
p}} \right)^2 \right] \ ,
\end{eqnarray}
with $\int_{\rm{\bf q}}$ being an abbreviation for $\left( 2 \pi \right)^{-d/2} \int d^d 
q$. It is convenient to switch back to continuous currents and replace the summation 
over the loop current by an integration,
\begin{eqnarray}
\mbox{A} &=& - \int_0^\infty ds_1 ds_2 D \left( {\rm{\bf p}}^2, \left\{ s_i \right\} 
\right) \int_{-\infty}^\infty d\stackrel{\mbox{{\tiny 
$\leftrightarrow$}}}{\kappa} \exp \left[ w P \left( \stackrel{\mbox{{\tiny 
$\leftrightarrow$}}}{\lambda} , \stackrel{\mbox{{\tiny 
$\leftrightarrow$}}}{\kappa} \right) \right] 
\nonumber \\
&\times & \Bigg\{  s_1 v_2 K_2 \left( \stackrel{\mbox{{\tiny 
$\leftrightarrow$}}}{\kappa} \right) + s_2  v_2 K_2 \left( 
\stackrel{\mbox{{\tiny $\leftrightarrow$}}}{\kappa} + \stackrel{\mbox{{\tiny 
$\leftrightarrow$}}}{\lambda} \right) \Bigg\} \ .
\end{eqnarray}
The integration over the loop current is simplified by completing the squares in the 
exponential. One looks for the minimum of the quadratic form $P\left( 
\stackrel{\mbox{{\tiny $\leftrightarrow$}}}{\lambda} , \stackrel{\mbox{{\tiny 
$\leftrightarrow$}}}{\kappa} \right)$. The minimum is determined by a variation principle completely analogous to the one stated in Eq.~(\ref{noisyVariationPrinciple2}). Thus 
completing the squares is equivalent to solving Kirchhoff's equations for the 
diagram. 
We obtain
\begin{eqnarray}
\mbox{A} &=& - \int_0^\infty ds_1 ds_2 D \left( {\rm{\bf p}}^2, \left\{ s_i \right\} 
\right) \exp \left[ -R \left(  \left\{ s_i \right\} \right) w \stackrel{\mbox{{\tiny 
$\leftrightarrow$}}}{\lambda}^2 \right] \int_{-\infty}^\infty d \stackrel{\mbox{{\tiny 
$\leftrightarrow$}}}{\kappa} \exp \left[ - \left( s_1 + s_2 \right) w 
\stackrel{\mbox{{\tiny $\leftrightarrow$}}}{\kappa}^2 \right] 
\nonumber \\
&\times & \Bigg\{ s_1 v_2 K_2 \left( \stackrel{\mbox{{\tiny 
$\leftrightarrow$}}}{\kappa} - \frac{s_2}{s_1 + s_2} \stackrel{\mbox{{\tiny 
$\leftrightarrow$}}}{\lambda} \right) + s_2 v_2 K_2 \left( \stackrel{\mbox{{\tiny 
$\leftrightarrow$}}}{\kappa} + \frac{s_1}{s_1 + s_2} 
\stackrel{\mbox{{\tiny $\leftrightarrow$}}}{\lambda} \right) \Bigg\} \ ,
\end{eqnarray}
where $R \left(  \left\{ s_i \right\} \right) = s_1 s_2 / (s_1 + s_2 )$ is the total 
resistance of the diagram. Note that $s_2 \stackrel{\mbox{{\tiny 
$\leftrightarrow$}}}{\lambda} /(s_1 + s_2)$ is, appart from a factor $i$, the replica 
current induced by the external replica current $\stackrel{\mbox{{\tiny 
$\leftrightarrow$}}}{\lambda}$ into the propagator parametrized by $s_1$. $s_1 
\stackrel{\mbox{{\tiny $\leftrightarrow$}}}{\lambda} /(s_1 + s_2)$ is the replica 
current induced into the propagator parametrized by $s_2$. In the limit $D \to 0$ we 
find
\begin{eqnarray}
\mbox{A} &=& - g^2 \frac{1}{\left( 4 \pi \right)^{d/2}} \int_0^\infty \frac{ds_1 
ds_2}{\left( s_1 + s_2 \right)^{d/2}} \exp \left[ - \left( s_1 + s_2 \right) \tau -R 
\left( 
 \left\{ s_i \right\} \right) \left( {\rm{\bf p}}^2 + w \stackrel{\mbox{{\tiny 
$\leftrightarrow$}}}{\lambda}^2 \right) \right]
\nonumber \\
&\times & \Bigg\{ \frac{s_1 s_2^4}{\left( s_1 + s_2 \right)^4} \, v_2 K_2 \left( 
\stackrel{\mbox{{\tiny $\leftrightarrow$}}}{\lambda} \right) + 2 \frac{s_1 s_2^2}{\left( 
s_1 + s_2 \right)^3} \, \frac{v_2}{w} \stackrel{\mbox{{\tiny 
$\leftrightarrow$}}}{\lambda}^2 \Bigg\} \ ,
\end{eqnarray}
where we have carried out the momentum integration as well. Expanding the exponential 
and 
keeping only the terms proportional to $v_2$ gives
\begin{eqnarray}
\mbox{A} &=& - g^2 \frac{1}{\left( 4 \pi \right)^{d/2}} \int_0^\infty \frac{ds_1 
ds_2}{\left( s_1 + s_2 \right)^{d/2}} \exp \left[ - \left( s_1 + s_2 \right) \tau 
\right]
\nonumber \\
&\times& \Bigg\{ \frac{s_1 s_2^4}{\left( s_1 + s_2 \right)^4} \, v_2 K_2 \left( 
\stackrel{\mbox{{\tiny $\leftrightarrow$}}}{\lambda} \right) + 2 \frac{s_1 s_2^2}{\left( 
s_1 + s_2 \right)^3} \, \frac{v_2}{w} \stackrel{\mbox{{\tiny 
$\leftrightarrow$}}}{\lambda}^2 
\nonumber \\
&-& 2 \frac{s_1^2 s_2^3}{\left( s_1 + s_2 \right)^4} \, \frac{v_2}{w} 
\stackrel{\mbox{{\tiny $\leftrightarrow$}}}{\lambda}^2 \left( {\rm{\bf p}}^2 + w 
\stackrel{\mbox{{\tiny $\leftrightarrow$}}}{\lambda}^2 \right) - \frac{s_1^2 
s_2^5}{\left( 
s_1 + s_2 \right)^5} \, v_2 K_2 \left( \stackrel{\mbox{{\tiny 
$\leftrightarrow$}}}{\lambda} 
\right) \left( {\rm{\bf p}}^2 + w \stackrel{\mbox{{\tiny $\leftrightarrow$}}}{\lambda}^2 
\right) \Bigg\} \ .
\end{eqnarray}
The integral over the last term is convergent and therefore neglected. The remaining 
integrations are rendered straightforward by the change of variables $s_1 \to tx$ and 
$s_2 
\to t \left( 1-x \right)$. Upon expanding the result for small $\epsilon = 6 -d$ we 
obtain
\begin{eqnarray}
\mbox{A} &=& - g^2 \frac{G_\epsilon}{\epsilon} \tau^{-\epsilon /2} \Bigg\{ \frac{1}{15} 
v_2 
K_2 \left( \stackrel{\mbox{{\tiny $\leftrightarrow$}}}{\lambda} \right) - \frac{1}{3} 
\frac{v_2}{w} \stackrel{\mbox{{\tiny $\leftrightarrow$}}}{\lambda}^2 \tau - \frac{1}{15} 
\frac{v_2}{w} \stackrel{\mbox{{\tiny $\leftrightarrow$}}}{\lambda}^2 \left( {\rm{\bf 
p}}^2 
+ w \stackrel{\mbox{{\tiny $\leftrightarrow$}}}{\lambda}^2 \right) \Bigg\} \ ,
\end{eqnarray}
where we have introduced $G_\epsilon = \left( 4 \pi \right)^{(-d/2)} \Gamma \left( 1 + 
\epsilon /2 \right)$ for convenience. We learn, that not only primitive divergencies proportional to $K_2 \left( \stackrel{\mbox{{\tiny $\leftrightarrow$}}}{\lambda} \right)$, but also proportional to $\tau \stackrel{\mbox{{\tiny $\leftrightarrow$}}}{\lambda}^2$, ${\rm{\bf p}}^2 \stackrel{\mbox{{\tiny $\leftrightarrow$}}}{\lambda}^2$ and $\left( \stackrel{\mbox{{\tiny $\leftrightarrow$}}}{\lambda}^2 \right)^2$ are generated.

\section{Superficial degree of divergence of operator insertions}
\label{app:superficialDegreeOfDivergenceOfOperatorInsertions}
Consider the insertion of a local operator $\mathcal O \mathnormal$ by adding a term to 
the Hamiltonian,
\begin{eqnarray}
\mathcal H \mathnormal \to \mathcal H \mathnormal + \int d^d x \ 
\sum_{\stackrel{\mbox{{\tiny 
$\leftrightarrow$}}}{\theta}} \mathcal O \mathnormal \left( {\bf{\rm x}}, 
\stackrel{\mbox{{\tiny $\leftrightarrow$}}}{\theta} \right) \ , 
\end{eqnarray} 
where $\mathcal O \mathnormal$ is a local monomial of degree $n$ in the fields $\varphi$ 
with $A$ derivatives in real and $B$ derivatives in replica space. In a diagram composed 
of $P$ propagators, $V$ three-leg vertices and the insertion there are
\begin{eqnarray}
\label{loopNumber}
L = P - \left( V+1 \right) - 1 
\end{eqnarray}
loops. The topological relation
\begin{eqnarray}
\label{topoRel}
3V + n = 2P + E 
\end{eqnarray}
balances the number of legs. Each propagator behaves for large momenta as $1/{\rm{\bf 
q}}^{2}$ and hence reduces the superficial degree of divergence of the diagram by 2. The 
insertion increases it by $A+B$. Thus the superficial degree of divergence $\delta 
\left[ 
\mathcal O \mathnormal \right]$ of the diagram with insertion is
\begin{eqnarray}
\delta \left[ \mathcal O \mathnormal \right] = dL +A + B - 2P \ .
\end{eqnarray}
With help of Eqs.~(\ref{loopNumber}) and (\ref{topoRel}) one finds
\begin{eqnarray}
\delta \left[ \mathcal O \mathnormal \right] = \frac{d-6}{2} V + \frac{d-2}{2} n + 
\frac{2-d}{2} E + A + B \ .
\end{eqnarray}
In contrast, the superficial degree of divergence $\delta$ of the diagram without 
insertion is
\begin{eqnarray}
\delta  = d + \frac{d-6}{2} V + \frac{2-d}{2} E \ .
\end{eqnarray}
The difference 
\begin{eqnarray}
\delta \left[ \mathcal O \mathnormal \right] - \delta = \frac{d-2}{2} n + A + B - d 
\end{eqnarray}
is identical to the naive dimension $\left[ \mathcal O \mathnormal \right]$ of the 
insertion. For $d=6$ it reduces to
\begin{eqnarray}
\label{naiveDimOfInsertion}
\left[ \mathcal O \mathnormal \right] = 2 n + A + B - d \ .
\end{eqnarray}

\section{Computation of Diagrams II}
\label{app:noisyComputationOfDiagrams}
Here we give details on the calculation of the conducting Feynman diagrams in 
listed in Fig.~1. We focus on the contributions of the diagrams to the 
renormalization of the $v_l$, i.e., those terms appearing in 
Eq.~(\ref{noisyExpansionOfDiagrams}) proportional to $v_l$. The other terms appearing in 
Eq.~(\ref{noisyExpansionOfDiagrams}) will be omitted throughout the entire appendix for 
the sake of notational simplicity. For details on the calculation of the contributions 
to 
the renormalization of $w$ we refer to\cite{stenull_janssen_oerding_99}. The 
$\stackrel{\mbox{{\tiny $\leftrightarrow$}}}{\lambda}$-independent parts of the 
conducting diagrams correspond to the usual diagrams found in the literature on the 
Potts 
model\cite{Zia_Wallace_75} and can be calculated by standard 
proceedures\cite{amit_zinn-justin}.

We start with diagram A. The part of A required in the calculation of $\psi_l$ reads for 
vanishing external momentum
\begin{eqnarray}
\label{exampleA2}
\mbox{A} = - \frac{g^2}{2}  v_l K_l \left( \stackrel{\mbox{{\tiny 
$\leftrightarrow$}}}{\lambda} \right) \int_0^\infty ds_1 ds_2 \int_{\rm{\bf q}} \exp 
\left[ - \left( s_1 + s_2 \right) \left( \tau + {\rm{\bf q}}^2 \right) \right] \bigg\{ 
s_1 
\left( \frac{s_2}{s_1 + s_2} \right)^{n} + s_2 \left( \frac{s_1}{s_1 + s_2} \right)^{n} 
\bigg\} \ ,
\end{eqnarray}
where $n=2l$. Carrying out the momentum integration gives
\begin{eqnarray}
\label{exampleA3}
\mbox{A} = - g^2  v_l K_l \left( \stackrel{\mbox{{\tiny $\leftrightarrow$}}}{\lambda} 
\right) \frac{1}{\left( 4 \pi \right)^{d/2}} \int_0^\infty ds_1 ds_2 \ \frac{1}{\left( 
s_1 
+ s_2 \right)^{d/2}} \exp \left[ - \left( s_1 + s_2 \right) \tau \right]  \frac{s_1 
s_2^n}{\left( s_1 + s_2 \right)^n}  \ .
\end{eqnarray}
Changing variables, $s_1 \to t \left( 1-x \right)$ and $s_2 \to tx$, leads to
\begin{eqnarray}
\label{exampleA4}
\mbox{A} &=& - g^2  v_l K_l \left( \stackrel{\mbox{{\tiny $\leftrightarrow$}}}{\lambda} 
\right) \frac{1}{\left( 4 \pi \right)^{d/2}} 
\int_0^1 dx \ \left( 1-x \right) x^n \int_0^\infty dt \ t^{2-d/2} \exp \left( - t \tau 
\right) 
\nonumber \\
&=& - g^2  v_l K_l \left( \stackrel{\mbox{{\tiny $\leftrightarrow$}}}{\lambda} 
\right) \frac{1}{\left( 4 \pi \right)^{d/2}} 
\frac{1}{\left( n+1 \right) \left( n+2 \right)} \Gamma \left( 3 - \frac{d}{2} \right) 
\tau^{d/2 -3} \ .
\end{eqnarray}
Expansion for small $\epsilon =6-d$ yields
\begin{eqnarray}
\label{exampleA6}
\mbox{A} = - g^2  v_l K_l \left( \stackrel{\mbox{{\tiny $\leftrightarrow$}}}{\lambda} 
\right) 
\frac{2}{\left( n+1 \right) \left( n+2 \right)} \frac{G_\epsilon}{\epsilon} 
\tau^{-\epsilon /2} \ ,
\end{eqnarray}
where we have introduced $G_\epsilon = \left( 4 \pi \right)^{-d/2} \Gamma \left( 1 + 
\epsilon /2 \right)$ for convenience.

The calculation of B is particularly simple. Thus we merely state the result
\begin{eqnarray}
\label{exampleB}
\mbox{B} = - g^2  v_l K_l \left( \stackrel{\mbox{{\tiny $\leftrightarrow$}}}{\lambda} 
\right) \frac{G_\epsilon}{2\epsilon} \tau^{-\epsilon /2} \ .
\end{eqnarray}

Now we turn to the two-loop diagrams. As an example, we consider the diagram C. As a 
first step, we determine the currents flowing through the conducting propagators. 
Kirchhoff's law Eq.~(\ref{noisyCirquitEquations}) applies to the 4 vertices of the 
diagram. This allows us to eliminate 3 of the 5 unknown currents (one of the vertices is 
inactive with respect to this purpose since the external current $\stackrel{\mbox{{\tiny 
$\leftrightarrow$}}}{\lambda}$ must be conserved). The potential drop around closed 
loops is zero. Hence we can eliminate the two remaining unknown currents and express all 
currents flowing through conducting propagators in terms of the Schwinger parameters and 
$\stackrel{\mbox{{\tiny $\leftrightarrow$}}}{\lambda}$. The momentum integrations are 
straightforward. They can be done by using the saddle point method which works exactly 
here since the momentum dependence is purely quadratic. After the momentum integration 
we 
have
\begin{eqnarray}
\label{exampleC1}
\mbox{C} &=& - \frac{g^4}{2}  v_l K_l \left( \stackrel{\mbox{{\tiny 
$\leftrightarrow$}}}{\lambda} \right) \frac{1}{\left( 4 \pi \right)^{d}} \int_0^\infty 
\prod_{i=1}^5 ds_i \  
\frac{\exp \left( - \tau \sum_{i=1}^5 s_i \right)}{\left[ \left( s_1 + s_2 + s_5 \right) 
\left( s_3 + s_4 + s_5 \right) - s_5^2 \right]^{d/2}}
\nonumber \\
&\times& \Bigg\{ s_1 \left[ \frac{s_2 \left( s_3 + s_4 + s_5 \right) + s_4 s_5}{\left( 
s_1 + s_2 + s_5 \right) \left( s_3 + s_4 + s_5 \right) - s_5^2 } \right]^n + s_2 \left[ 
\frac{s_1 \left( s_3 + s_4 + s_5 \right) + s_3 s_5}{\left( s_1 + s_2 + s_5 \right) 
\left( 
s_3 + s_4 + s_5 \right) - s_5^2 } \right]^n
\nonumber \\
& & + s_3 \left[ \frac{s_4 \left( s_3 + s_4 + s_5 \right) + s_2 s_5}{\left( s_1 + s_2 + 
s_5 \right) \left( s_3 + s_4 + s_5 \right) - s_5^2 } \right]^n + s_4 \left[ \frac{s_3 
\left( s_3 + s_4 + s_5 \right) + s_1 s_5}{\left( s_1 + s_2 + s_5 \right) \left( s_3 + 
s_4 
+ s_5 \right) - s_5^2 } \right]^n
\nonumber \\
& & + s_5 \left[ \frac{s_2 s_3 - s_1 s_4}{\left( s_1 + s_2 + s_5 \right) \left( s_3 + 
s_4 
+ s_5 \right) - s_5^2 } \right]^n \Bigg\}
\nonumber \\
&=& - \frac{g^4}{2}  v_l K_l \left( \stackrel{\mbox{{\tiny $\leftrightarrow$}}}{\lambda} 
\right) \frac{1}{\left( 4 \pi \right)^{d}} \int_0^\infty \prod_{i=1}^5 ds_i \  
\frac{\exp \left( - \tau \sum_{i=1}^5 s_i \right)}{\left[ \left( s_1 + s_2 + s_5 \right) 
\left( s_3 + s_4 + s_5 \right) - s_5^2 \right]^{d/2+n}}
\nonumber \\
&\times& \Bigg\{ 4 s_1 \left[ s_2 \left( s_3 + s_4 + s_5 \right) + s_4 s_5 \right]^n  + 
s_5 \left[ s_2 s_3 - s_1 s_4 \right]^n \Bigg\} \ .
\end{eqnarray} 
At this stage, the change of variables $s_1 \to t_1 \left( 1-x \right)$, $s_2 \to t_1 
x$, $s_3 \to t_2 \left( 1-y \right)$, $s_4 \to t_2 y$, and $s_5 \to t_3$ turns out to be 
useful. It leads to
\begin{eqnarray}
\label{exampleC2}
\mbox{C} &=&  - \frac{g^4}{2}  v_l K_l \left( \stackrel{\mbox{{\tiny 
$\leftrightarrow$}}}{\lambda} \right) \frac{1}{\left( 4 \pi \right)^{d}} \int_0^\infty 
dt_1 dt_2 dt_3 \int_0^1 dx dy \  \frac{\exp \left[ - \tau \left( t_1 + t_2 + t_3  
\right) 
\right]}{\left[ t_1 t_2 + t_2 t_3 + t_1 t_3 \right]^{d/2+n}} \ t_1 t_2
\nonumber \\
&\times& \Bigg\{ 4 t_1 \left( 1-x \right) \left[ x t_1 \left( t_2 + t_3 \right) + y t_2 
t_3 \right]^n  + t_3 \left( t_1 t_2 \right)^n \left[ x-y \right]^n \Bigg\} \ .
\end{eqnarray} 
The integrations over $x$ and $y$ are straightforward and can be basically looked up in 
a table\cite{gradshteyn_ryzhik}. After some additional algebra we obtain
\begin{eqnarray}
\label{exampleC3}
\mbox{C} &=&  - \frac{g^4}{2}  v_l K_l \left( \stackrel{\mbox{{\tiny 
$\leftrightarrow$}}}{\lambda} \right) \frac{1}{\left( 4 \pi \right)^{d}} \int_0^\infty 
dt_1 dt_2 dt_3 \ \frac{\exp \left[ - \tau \left( t_1 + t_2 + t_3  \right) 
\right]}{\left[ 
t_1 t_2 + t_2 t_3 + t_1 t_3 \right]^{d/2}}
\nonumber \\
&\times& \Bigg\{ 
\frac{2}{\left( n+1 \right) \left( n+2 \right)} \ 
\frac{t_3 \left( t_1 t_2 \right)^{n+1}}{\left[ t_1 t_2 + t_2 t_3 + t_1 t_3 \right]^{n}}
\nonumber \\
& & + \frac{4}{\left( n+1 \right) \left( n+2 \right) \left( n+3 \right)} \ \frac{t_3 
\left[ t_1 t_2 + t_2 t_3 + t_1 t_3 \right]^2}{t_1 \left( t_1 + t_2\right)} 
\nonumber \\
& &+ \frac{4}{\left( n+1 \right) \left( n+2 \right) \left( n+3 \right)} \ 
\frac{t_1 \left[ t_1 t_2 + t_2 t_3 + t_1 t_3 \right]^2}{\left( t_1 + t_2 \right)^2}
\nonumber \\
& & - \frac{4}{\left( n+1 \right) \left( n+2 \right) \left( n+3 \right)} \ 
\frac{t_3^{n+3} 
\left( t_1 + t_2 \right)^{n+1}}{t_1 \left[ t_1 t_2 + t_2 t_3 + t_1 t_3 \right]^{n}} 
\nonumber \\
& &- \frac{4}{\left( n+1 \right) \left( n+2 \right) \left( n+3 \right)} \ 
\frac{t_1^{n+3} 
t_2^{n+2}}{\left(t_1 + t_2 \right)^2 \left[ t_1 t_2 + t_2 t_3 + t_1 t_3 \right]^{n}} 
\nonumber \\
& & - \frac{4}{\left( n+1 \right) \left( n+2 \right)} \
\frac{t_3 t_1^{n+2} t_2^{n+1}}{\left(t_1 + t_2 \right) \left[ t_1 t_2 + t_2 t_3 + t_1 
t_3 
\right]^{n}}
\Bigg\} \ ,
\end{eqnarray} 
We find it convenient to express the remaining integrals in terms of the parameter 
integrals given in App.~\ref{app:parameterIntegrals}. For the sake of notational 
simplicity we introduce the notation
\begin{eqnarray}
\label{notationM1}
M_{i,j,k}^\mu = \left( -1 \right)^{i+j+k} \frac{\partial^{i+j+k}}{\partial a^i \partial 
b^j \partial c^k} M^\mu \left( a, b, c \right) \bigg|_{a=b=c=\tau}\ ,
\end{eqnarray}
where $\mu \in \left\{ 1, 3, 4\right\}$, and
\begin{eqnarray}
\label{notationM2}
M_{i,j,k}^\mu \left( n \right) = \left( -1 \right)^{i+j+k} 
\frac{\partial^{i+j+k}}{\partial a^i \partial b^j \partial c^k} M^\mu \left( a, b, c; n 
\right) \bigg|_{a=b=c=\tau}\ ,
\end{eqnarray}
where $\mu \in \left\{ 5,6,7,8 \right\}$. In terms of the parameter integrals we obtain
\begin{eqnarray}
\label{exampleC4}
\mbox{C} &=&  - \frac{g^4}{2}  v_l K_l \left( \stackrel{\mbox{{\tiny 
$\leftrightarrow$}}}{\lambda} \right) \Bigg\{ \frac{2}{\left( n+1 \right) \left( n+2 
\right)} \bigg[ M_{1,0,0}^8 \left( n \right) - 2 M_{2,0,1}^7 \left( n \right) -4 
M_{1,1,1}^7 \left( n \right) - 2 M_{0,2,1}^7 \left( n \right) \bigg]
\nonumber \\
&+& \frac{4}{\left( n+1 \right) \left( n+2 \right) \left( n+3 \right)} \bigg[ M + 
M_{2,0,0}^3 + M_{1,1,0}^3 
\nonumber \\
&-&  M_{1,0,0}^6 \left( n \right) -  M_{0,1,0}^6 \left( n \right) -  M_{3,0,0}^7 \left( 
n \right) -  M_{2,1,0}^7 \left( n \right) \bigg] 
\Bigg\} \ .
\end{eqnarray} 
The final result reads
\begin{eqnarray}
\label{exampleC5}
\mbox{C} &=&  - \frac{g^4}{2}  v_l K_l \left( \stackrel{\mbox{{\tiny 
$\leftrightarrow$}}}{\lambda} \right) \frac{G_\epsilon^2}{\epsilon} \tau^{-\epsilon} 
\Bigg\{ \frac{4n + 12}{\left( n+1 \right) \left( n+2 \right) \left( n+3 \right) 
\epsilon}
\nonumber \\
&+& \frac{4n - 2 F_2 \left( n+3 \right) +12}{\left( n+1 \right) \left( n+2 \right) 
\left( 
n+3 \right)} - \frac{2}{\left( n+1 \right) \left( n+2 \right)^2 \left( n+3 \right)}
\Bigg\} \ .
\end{eqnarray}
The diagrams D to G can be evaluated in the same fashion. 

As another example we consider diagram H. Determination of the noise cumulants of H 
leads 
to
\begin{eqnarray}
\label{exampleH1}
\mbox{H} &=& - \frac{g^4}{2}  v_l K_l \left( \stackrel{\mbox{{\tiny 
$\leftrightarrow$}}}{\lambda} \right) \frac{1}{\left( 4 \pi \right)^{d}} \int_0^\infty 
\prod_{i=1}^5 ds_i \  
\frac{\exp \left( - \tau \sum_{i=1}^5 s_i \right)}{\left[ \left( s_1 + s_2 + s_5 \right) 
\left( s_3 + s_4 \right) + s_3 s_4 \right]^{d/2+n}}
\nonumber \\
&\times& \Bigg\{ 2 s_1 \left[ s_5 \left( s_3 + s_4 \right) \right]^n + 2 s_3 \left[s_4 
s_5 
 \right]^n + s_5 \left[ \left( s_1 + s_2 \right) \left( s_3 + s_4 \right) + s_3 s_4 
\right]^n \Bigg\} \ .
\end{eqnarray}
Here, the change of variables $s_5 \to t_1 x$, $s_1 \to t_1 y$, $s_2 \to t_1 \left( 
1-x-y \right)$, $s_3 \to t_2$, and $s_4 \to t_3$ simplifies the integration. We obtain
\begin{eqnarray}
\label{exampleH2}
\mbox{H} &=&  - \frac{g^4}{2}  v_l K_l \left( \stackrel{\mbox{{\tiny 
$\leftrightarrow$}}}{\lambda} \right) \frac{1}{\left( 4 \pi \right)^{d}} \int_0^\infty 
dt_1 dt_2 dt_3 \int_0^1 dx \int_0^{1-x} dy \ \frac{\exp \left[ - \tau \left( t_1 + t_2 + 
t_3  \right) \right]}{\left[ t_1 t_2 + t_2 t_3 + t_1 t_3 \right]^{d/2+n}} \ t_1^2
\nonumber \\
&\times& \Bigg\{ 2 y t_1 \left[ x t_1  \left( t_2 + t_3 \right) \right]^n + 2 t_2 \left[ 
x 
t_1 t_3  \right]^n + x t_1 \left[ \left( 1-x \right) t_1 \left( t_2 + t_3 \right)  + t_2 
t_3 \right]^n \Bigg\} \ .
\end{eqnarray} 
The integrations over $x$ and $y$ are again straightforward. In terms of the parameter 
integrals we find
\begin{eqnarray}
\label{exampleH3}
\mbox{H} &=&  - \frac{g^4}{2}  v_l K_l \left( \stackrel{\mbox{{\tiny 
$\leftrightarrow$}}}{\lambda} \right) \Bigg\{ 
- \frac{1}{\left( n+1 \right)} \bigg[ M_{1,0,1}^4 + M_{0,1,1}^4 + M_{1,1,0}^4 \bigg]
\nonumber \\
&+& \frac{1}{\left( n+1 \right) \left( n+2 \right)} \bigg[ 2 M_{0,1,0}^8 \left( n 
\right) 
+ M_{2,0,1}^7 \left( n \right) + M_{1,1,1}^7 \left( n \right) \bigg]
\nonumber \\
&+& \frac{1}{\left( n+2 \right) \left( n+3 \right)} \bigg[ M_{1,0,1}^3 + M_{0,1,1}^3 + 
\left( n+4 \right) M_{1,1,0}^3 \bigg]
\nonumber \\
&+& \frac{2}{\left( n+1 \right) \left( n+2 \right) \left( n+3 \right)} \bigg[ 
M_{1,0,0}^6 
\left( n \right) + M_{2,1,0}^7 \left( n \right) \bigg] 
\Bigg\} \ .
\end{eqnarray} 
Finally we obtain
\begin{eqnarray}
\label{exampleH4}
\mbox{H} &=&  - \frac{g^4}{2}  v_l K_l \left( \stackrel{\mbox{{\tiny 
$\leftrightarrow$}}}{\lambda} \right) \frac{G_\epsilon^2}{\epsilon} \tau^{-\epsilon} 
\Bigg\{ - \frac{1}{\left( n+1 \right) \left( n+2 \right) \epsilon} + \frac{4}{\left( n+1 
\right)^2 \left( n+2 \right)^2 \epsilon}
\nonumber \\
&-& \frac{8n - F_2 \left( n+3 \right) + 24}{3 \left( n+1 \right) \left( n+2 \right) 
\left( 
n+3 \right)} - \frac{11}{6 \left( n+1 \right)^2 \left( n+2 \right)^2} 
\nonumber \\
&+& \frac{1}{3 \left( n+1 \right) \left( n+2 \right)^2 \left( n+3 \right)}
+ \frac{4n + 6}{\left( n+1 \right)^3 \left( n+2 \right)^3} \Bigg\} \ .
\end{eqnarray}
The diagrams I to L can be treated in a similar manner. 

\section{Parameter Integrals}
\label{app:parameterIntegrals}
This appendix contains a list of the parameter integrals we use in the calculation of 
the 
noise exponents. The results stated are obtained by employing the dimensional 
regularization scheme. The parameter integral $M^1$ given below was introduced by Breuer 
and Janssen\cite{breuer_janssen_81}. The notation $M^2$ we reserved for a parameter 
integral we introduced in\cite{stenull_janssen_oerding_99} but which is not used here.  
For notational briefness we define 
\begin{eqnarray}
F_m \left( n \right) = \sum_{k=m}^n {n \choose k}(-1)^k \frac{1}{k-m+1}
\end{eqnarray}
$F_m$ is related to the Digamma function $\Psi$ via
\begin{eqnarray}
F_m \left( n \right) = (-1)^{m+1} \frac{n! \left[ \Psi (m) - \Psi (n+1) 
\right]}{(m-1)!\, 
(1-m+n)!} \ .
\end{eqnarray}

The parameter integrals we use in calculating the $\psi_l$ are:
\begin{eqnarray}
\label{M1}
\lefteqn{ M^1 \left( a, b, c \right)=
\int_{{\rm{\bf p}} ,{\rm{\bf q}}} \frac{1}{ \left( a + {\rm{\bf p}}^2 \right) \left( b + 
{\rm{\bf q}}^2 \right) \left( c + \left( {\rm{\bf p}} + {\rm{\bf q}} \right)^2 \right)} 
}
\nonumber \\
&=& \frac{1}{(4\pi)^d} \int_0^\infty dt_1 dt_2 dt_3 \
\frac{\exp \left[ - \left( at_1 + bt_2 + ct_3 \right) \right] }{\left[ t_3 t_1 + t_3 t_2 
+ 
t_1 t_2 \right]^{d/2} } 
\nonumber \\
&=& \frac{G_\epsilon^2}{6\epsilon} \left\{ \left( \frac{1}{\epsilon} + \frac{25}{12} 
\right) \left( a^{3-\epsilon} + b^{3-\epsilon} + c^{3-\epsilon} \right) \right.
\nonumber \\
&-& \left. \left( \frac{3}{\epsilon} + \frac{21}{4} \right) \left[ a^{2-\epsilon} \left( 
b 
+ c \right) + b^{2-\epsilon} \left( a + c \right) + c^{2-\epsilon} \left( a + b \right) 
\right] - 3abc \right\} \ ,
\end{eqnarray}
\begin{eqnarray}
\label{M3}
\lefteqn{ M^3 \left( a, b, c \right) = \frac{1}{(4\pi)^d} \int_0^\infty dt_1 dt_2 dt_3 \ 
\frac{\exp \left[ - \left( at_1 + bt_2 + ct_3 \right) \right] }{\left[ t_3 t_1 + t_3 t_2 
+ 
t_1 t_2 \right]^{d/2-2} } \ \frac{1}{\left[ t_1 + t_2 \right]^3} }
\nonumber \\
&=& \frac{G_\epsilon^2}{2 \epsilon} \bigg\{ c^{2-\epsilon} \left( \frac{1}{15\epsilon} + 
\frac{46}{450} \right) - c^{1-\epsilon} \left( a+b \right) \left( \frac{1}{3\epsilon} + 
\frac{4}{9} \right)
\nonumber \\
&+& c^{-\epsilon} \left( a^2+b^2 \right) \left( \frac{2}{3\epsilon} + \frac{13}{18} 
\right) + c^{-\epsilon} ab \left( \frac{2}{3\epsilon} + \frac{5}{9} \right) \bigg\} \ ,
\end{eqnarray}
\begin{eqnarray}
\label{M4}
\lefteqn{ M^4 \left( a, b, c \right) = \frac{1}{(4\pi)^d} \int_0^\infty dt_1 dt_2 dt_3 \ 
\frac{\exp \left[ - \left( at_1 + bt_2 + ct_3 \right) \right] }{\left[ t_3 t_1 + t_3 t_2 
+ 
t_1 t_2 \right]^{d/2-1} } \ \frac{t_1t_2}{\left[ t_1 + t_2 \right]^3} }
\nonumber \\
&=& \frac{G_\epsilon^2}{2 \epsilon} \bigg\{ - c^{2-\epsilon} \left( \frac{2}{15\epsilon} 
+ 
\frac{107}{450} \right) - c^{1-\epsilon} \left( a+b \right) \left( \frac{1}{3\epsilon} + 
\frac{4}{9} \right)
\nonumber \\
&+& \frac{1}{3} \left( a^2+b^2 \right) + \frac{1}{3} ab \bigg\} \ ,
\end{eqnarray}
\begin{eqnarray}
\label{M5}
\lefteqn{ M^5 \left( a, b, c; n \right) = \frac{1}{(4\pi)^d} \int_0^\infty dt_1 dt_2 
dt_3 
\ \frac{\exp \left[ - \left( at_1 + bt_2 + ct_3 \right) \right] }{\left[ t_3 t_1 + t_3 
t_2 
+ t_1 t_2 \right]^{d/2} } \ \frac{t_1^n}{\left[ t_1 + t_2 \right]^n} }
\nonumber \\
&=& - \frac{G_\epsilon^2}{6\epsilon} \bigg\{ a^{3-\epsilon} \left[ - 
\frac{n+1}{\epsilon} 
- \frac{25 \left( n+1 \right)}{12} + \frac{1}{2} F_2 \left( n+1 \right) \right]
\nonumber \\
&+& c^{3-\epsilon} \frac{1}{\left( n+2 \right) \left( n+3 \right)} \left[ - 
\frac{6}{\epsilon} - \frac{15}{2} + 3 \Psi \left( n+2 \right) + 3 \Psi \left( 2 \right) 
- 
6 \Psi \left( n+4 \right) \right] 
\nonumber \\
&+& a^{2-\epsilon} b \left[ \frac{3}{\epsilon} + \frac{21}{4} + \frac{3}{2} F_1 \left( n 
\right) \right] + a^{2-\epsilon} c \left[ \frac{3}{\epsilon} + \frac{27}{4} + 
\frac{3}{2} 
F_1 \left( n+1 \right) \right]
\nonumber \\
&+& c^{2-\epsilon} a \frac{1}{n+2} \left[ \frac{6}{\epsilon} + \frac{9}{2} - 3 \Psi 
\left( 
n+2 \right) + 6 \Psi \left( n+3 \right) + 3\gamma \right] 
\nonumber \\
&+& c^{2-\epsilon} b \frac{1}{\left( n+1 \right) \left( n+2 \right)} \left[ 
\frac{6}{\epsilon} + \frac{9}{2} - 3 \Psi \left( n+1 \right) - 3 \Psi \left( 2 \right) + 
6 
\Psi \left( n+3 \right) \right] 
\nonumber \\
&+& b^3 \frac{1}{2 \left( n-1 \right) n} + b^2 a \frac{3}{2n} + b^2 c  \frac{3}{2n 
\left( 
n+1 \right)} - 3abc \frac{1}{n+1} \bigg\} \ ,
\end{eqnarray}
\begin{eqnarray}
\label{M6}
\lefteqn{ M^6 \left( a, b, c; n \right) = \frac{1}{(4\pi)^d} \int_0^\infty dt_1 dt_2 
dt_3 
\ \frac{\exp \left[ - \left( at_1 + bt_2 + ct_3 \right) \right] }{\left[ t_3 t_1 + t_3 
t_2 
+ t_1 t_2 \right]^{d/2+n} } \ \frac{t_3^n \left[ t_1 + t_2 \right]^n}{t_1} }
\nonumber \\
&=& - \frac{G_\epsilon^2}{\epsilon} \bigg\{ c^{1-\epsilon} \left[ \frac{\left( n+2 
\right) 
\left( n+3 \right) }{12\epsilon} + \frac{n+3}{6 \epsilon}  + \frac{11 \left( n+2 \right) 
\left( n+3 \right) }{72} + \frac{2 \left( n+3 \right) }{9} - \frac{1}{24} \right.
\nonumber \\ 
&+& \left. \frac{1}{12} \left( F_3 \left( n+3 \right) - F_2 \left( n+3 \right)\right) 
\right]
\nonumber \\
&+& c^{-\epsilon} a  \left[ - \frac{n+3}{3 \epsilon} - \frac{5 \left( n+3 \right) }{18} 
+ 
\frac{1}{6} + \frac{1}{6} F_2 \left( n+3 \right) \right]
\nonumber \\
&+& c^{-\epsilon} b  \left[ - \frac{2 \left( n+3 \right)}{3 \epsilon} - \frac{13 \left( 
n+3 \right) }{18} + \frac{1}{6} + \frac{1}{3} F_2 \left( n+3 \right) \right]
\nonumber \\
&+& c^{-(1+\epsilon )} b^2 \left( \frac{2}{\epsilon} + \frac{3}{2} \right) - \frac{1}{2} 
c^{-(1+\epsilon )} a^2 - c^{-(1+\epsilon )} ab \bigg\} \ ,
\end{eqnarray}
\begin{eqnarray}
\label{M7}
\lefteqn{ M^7 \left( a, b, c; n \right) = \frac{1}{(4\pi)^d} \int_0^\infty dt_1 dt_2 
dt_3 
\ \frac{\exp \left[ - \left( at_1 + bt_2 + ct_3 \right) \right] }{\left[ t_3 t_1 + t_3 
t_2 
+ t_1 t_2 \right]^{d/2+n} } \ \frac{t_1^{n+1}t_2^{n+2}}{\left[ t_1+t_2 \right]^3} }
\nonumber \\
&=& - \frac{G_\epsilon^2}{6\epsilon} \bigg\{ b^{3-\epsilon} \left[ \frac{2}{\left( n+2 
\right) \epsilon} + \frac{1}{\left( n+2 \right)^2}  + \frac{11}{6 \left( n+2 \right)} 
\right]+ a^3 \frac{1}{3 \left( n+2 \right)} 
\nonumber \\
&+& c^3 \frac{1}{10 \left( n-1 \right) n\left( n+1 \right) \left( n+2 \right)} + a^2 b 
\frac{1}{2 \left( n+2 \right)} + a^2 c \frac{1}{4 \left( n+1 \right) \left( n+2 \right)}
\nonumber \\
&+& b^2 a \frac{1}{n+2} + b^2 c \frac{3}{4 \left( n+1 \right) \left( n+2 \right)} + c^2 
a 
\frac{1}{5 n \left( n+1 \right) \left( n+2 \right)}
\nonumber \\
&+& c^2 b \frac{3}{10 n \left( n+1 \right) \left( n+2 \right)} + abc \frac{1}{2 \left( 
n+1 
\right) \left( n+2 \right)} \bigg\} \ ,
\end{eqnarray}
\begin{eqnarray}
\label{M8}
\lefteqn{ M^8 \left( a, b, c; n \right) = \frac{1}{(4\pi)^d} \int_0^\infty dt_1 dt_2 
dt_3 
\ \frac{\exp \left[ - \left( at_1 + bt_2 + ct_3 \right) \right] }{\left[ t_3 t_1 + t_3 
t_2 
+ t_1 t_2 \right]^{d/2+n} } \ t_1^{n}t_2^{n+1}t_3 }
\nonumber \\
&=& - \frac{G_\epsilon^2}{\epsilon} \bigg\{ b^{1-\epsilon} \left[ \frac{2}{\left( n+1 
\right) \left( n+2 \right) \epsilon} + \frac{2n+3}{\left( n+1 \right)^2 \left( n+2 
\right)^2}  + \frac{1}{\left( n+1 \right) \left( n+2 \right)} \right]  
\nonumber \\
&+& a \frac{1}{\left( n+1 \right) \left( n+2 \right)} + c \frac{1}{ n \left( n+1 \right) 
\left( n+2 \right)} \bigg\} \ .
\end{eqnarray}

In addition to the parameter integrals we use
\begin{eqnarray}
\label{M}
M &=& \frac{1}{(4\pi)^d} \int_0^\infty dt_1 dt_2 dt_3 \
\frac{\exp \left[ - \tau \left( t_1 + t_2 + t_3 \right) \right] }{\left[ t_3 t_1 + t_3 
t_2 
+ t_1 t_2 \right]^{d/2-2} } \ \frac{t_3}{t_1 \left( t_1 + t_2 \right)} 
\nonumber \\
&=& \frac{G_\epsilon^2}{\epsilon} \tau^{-\epsilon} \left( \frac{3}{\epsilon} - 
\frac{1}{2} 
\right) \ .
\end{eqnarray}

\section{Conducting Diagrams in Terms of Parameter Integrals}
\label{app:diagramsInParameterIntegrals}
Here we list our results for the conducting two-loop diagrams in terms of the parameter 
integrals given in App.~\ref{app:parameterIntegrals}. For notational simplicity, we show 
only the parts of the diagrams proportional to $v_l K_l \left( \stackrel{\mbox{{\tiny 
$\leftrightarrow$}}}{\lambda} \right)$:
\begin{eqnarray}
\label{listC}
\mbox{C} &=&  - \frac{g^4}{2}  v_l K_l \left( \stackrel{\mbox{{\tiny 
$\leftrightarrow$}}}{\lambda} \right) \Bigg\{ \frac{2}{\left( n+1 \right) \left( n+2 
\right)} \bigg[ M_{1,0,0}^8 \left( n \right) - 2 M_{2,0,1}^7 \left( n \right) -4 
M_{1,1,1}^7 \left( n \right) - 2 M_{0,2,1}^7 \left( n \right) \bigg]
\nonumber \\
&+& \frac{4}{\left( n+1 \right) \left( n+2 \right) \left( n+3 \right)} \bigg[ M + 
M_{2,0,0}^3 + M_{1,1,0}^3 
\nonumber \\
&-&  M_{1,0,0}^6 \left( n \right) -  M_{0,1,0}^6 \left( n \right) -  M_{3,0,0}^7 \left( 
n 
\right) -  M_{2,1,0}^7 \left( n \right) \bigg] 
\Bigg\} \ ,
\end{eqnarray}
\begin{eqnarray}
\label{listD}
\mbox{D} &=&  - \frac{g^4}{2}  v_l K_l \left( \stackrel{\mbox{{\tiny 
$\leftrightarrow$}}}{\lambda} \right) \Bigg\{ \frac{n^2 + 3n + 6}{2 \left( n+1 \right) 
\left( n+2 \right)} M_{2,1,0}^1 + \frac{2}{\left( n+1 \right) \left( n+2 \right)} 
M_{1,1,1}^1 \Bigg\} \ ,
\end{eqnarray}
\begin{eqnarray}
\label{listE}
\mbox{E} &=&  - \frac{g^4}{2}  v_l K_l \left( \stackrel{\mbox{{\tiny 
$\leftrightarrow$}}}{\lambda} \right) \Bigg\{ - \frac{4}{\left( n+1 \right) \left( n+2 
\right)} M_{2,1,0}^5 \left( n \right)
\nonumber \\
&+& \frac{8}{\left( n+1 \right) \left( n+2 \right) \left( n+3 \right)} \bigg[ 
M_{3,0,0}^1 
+ 3 M_{2,1,0}^1 - M_{3,0,0}^5 \left( n \right) \bigg] \Bigg\} \ ,
\end{eqnarray}
\begin{eqnarray}
\label{listF}
\mbox{F} &=&  - \frac{g^4}{2}  v_l K_l \left( \stackrel{\mbox{{\tiny 
$\leftrightarrow$}}}{\lambda} \right) \Bigg\{ M_{2,1,0}^1 + M_{1,1,1}^1 \Bigg\} \ ,
\end{eqnarray}
\begin{eqnarray}
\label{listG}
\mbox{G} &=&  - \frac{g^4}{2}  v_l K_l \left( \stackrel{\mbox{{\tiny 
$\leftrightarrow$}}}{\lambda} \right) M_{2,1,0}^1 \ ,
\end{eqnarray}
\begin{eqnarray}
\label{listH}
\mbox{H} &=&  - \frac{g^4}{2}  v_l K_l \left( \stackrel{\mbox{{\tiny 
$\leftrightarrow$}}}{\lambda} \right) \Bigg\{ 
- \frac{1}{\left( n+1 \right)} \bigg[ M_{1,0,1}^4 + M_{0,1,1}^4 + M_{1,1,0}^4 \bigg]
\nonumber \\
&+& \frac{1}{\left( n+1 \right) \left( n+2 \right)} \bigg[ 2 M_{0,1,0}^8 \left( n 
\right) 
+ M_{2,0,1}^7 \left( n \right) + M_{1,1,1}^7 \left( n \right) \bigg]
\nonumber \\
&+& \frac{1}{\left( n+2 \right) \left( n+3 \right)} \bigg[ M_{1,0,1}^3 + M_{0,1,1}^3 + 
\left( n+4 \right) M_{1,1,0}^3 \bigg]
\nonumber \\
&+& \frac{2}{\left( n+1 \right) \left( n+2 \right) \left( n+3 \right)} \bigg[ 
M_{1,0,0}^6 
\left( n \right) + M_{2,1,0}^7 \left( n \right) \bigg] 
\Bigg\} \ ,
\end{eqnarray}
\begin{eqnarray}
\label{listI}
\mbox{I} &=&  - \frac{g^4}{2}  v_l K_l \left( \stackrel{\mbox{{\tiny 
$\leftrightarrow$}}}{\lambda} \right) \Bigg\{ 
\frac{1}{3} M_{3,0,0}^1 + M_{0,1,2}^5 \left( n \right) \Bigg\} \ ,
\end{eqnarray}
\begin{eqnarray}
\label{listJ}
\mbox{J} &=&  - \frac{g^4}{2}  v_l K_l \left( \stackrel{\mbox{{\tiny 
$\leftrightarrow$}}}{\lambda} \right) \Bigg\{ 
\frac{2}{\left( n+1 \right) \left( n+2 \right)} M_{2,1,0}^5 \left( n \right) 
\nonumber \\
&+& \frac{1}{\left( n+1 \right) \left( n+2 \right) \left( n+3 \right)} \bigg[ 4 
M_{3,0,0}^5 \left( n \right) + \left( n-1 \right) M_{3,0,0}^1 + 3 \left( n-1 \right) 
M_{2,1,0}^1 \bigg] 
\Bigg\} \ ,
\end{eqnarray}
\begin{eqnarray}
\label{listK}
\mbox{K} &=&  - \frac{g^4}{2}  v_l K_l \left( \stackrel{\mbox{{\tiny 
$\leftrightarrow$}}}{\lambda} \right) \Bigg\{ \frac{1}{3} M_{3,0,0}^1 + \frac{1}{2} 
M_{2,1,0}^1
\Bigg\} \ ,
\end{eqnarray}
\begin{eqnarray}
\label{listL}
\mbox{L} &=&  - \frac{g^4}{2}  v_l K_l \left( \stackrel{\mbox{{\tiny 
$\leftrightarrow$}}}{\lambda} \right) \frac{1}{6} M_{3,0,0}^1 \ .
\end{eqnarray}

\section{Relation to the backbone and the red bond dimension}
\label{app:relationToTheBackboneDimension}
From Eq.~(\ref{finalCumulant}) it is evident that only those bonds with $I_b = I$ 
contribute to $C_R^{(\infty)}$. Consequently, $\psi_\infty$ is related to the fractal 
dimension $d_{\mbox{{\scriptsize red}}}$ of the singly connected (red) bonds via 
$d_{\mbox{{\scriptsize red}}} = \psi_\infty/\nu$. Coniglio   
\cite{coniglio_81,coniglio_82} proved that $d_{\mbox{{\scriptsize red}}} = 1/\nu$, which 
in turn leads to $\psi_\infty =1$. As mentioned above, our result for $\psi_l$ matches 
this consistency requirement.

Another trivial consequence of Eq.~(\ref{finalCumulant}) is, that $C_R^{(0)}$ is proportional to the average number of bonds (the mass) of the backbone. Hence $\psi_0$ is related to the backbone dimension $D_B$ by
\begin{eqnarray}
\label{psiNullDB}
\psi_0 = \nu D_B \ .
\end{eqnarray}  
This relation can also be obtained on the level of Feynman diagrams. Revisit the definition of the noise cumulants for Feynman diagrams 
Eq.~(\ref{diagramCumulant}). In the limit $l \to 0$ the noise cumulant reduces to the 
sum of Schwinger parameters of conducting propagators,
\begin{eqnarray}
C^{(0)} \left(  \left\{ s_i \right\} \right) = \sum_i s_i  \ .
\end{eqnarray}
Now we take a short detour to our renormalized field theory of diluted networks in which 
the occupied bond obey a generalized Ohm's law $V\sim 
I^r$\cite{janssen_stenull_oerding_99,janssen_stenull_99}. In these networks, the 
nonlinear resistance $R_r \left( x, x^\prime \right)$ averaged subject to $x$ and 
$x^\prime$ being on the same cluster,
\begin{eqnarray}
M_r (x ,x^\prime) = \left\langle \chi (x ,x^\prime) R_r (x ,x^\prime )  \right\rangle_C 
/ \left\langle \chi (x ,x^\prime) \right\rangle_C \ , 
\end{eqnarray}
obeys at criticality
\begin{eqnarray}
M_r (x ,x^\prime) = \left| x - x^\prime \right|^{\phi_r/\nu} \ .
\end{eqnarray}
In analogy to $R \left( \left\{ s_i \right\} \right)$ we introduced the notion of the 
nonlinear resistance $R_r \left( \left\{ s_i \right\} \right)$ of Feynman diagrams. In 
the limit $r \to -1^+$ we found 
\begin{eqnarray}
R_{-1} \left( \left\{ s_i \right\} \right) = \sum_i s_i  \ .
\end{eqnarray}
Hence we can identify $C^{(0)} \left(  \left\{ s_i \right\} \right)$ and $R_{-1} \left( 
\left\{ s_i \right\} \right)$. This leads to the conclusion that $\psi_0 = \phi_{-1}$. 
$\phi_{-1}$ is related to the fractal dimension $D_B$ of the backbone via $\phi_{-1} = 
\nu D_B$, and hence we obtain once more Eq.~(\ref{psiNullDB}). Equation~(\ref{psiNullDB}) provides another consistency check for our result 
(\ref{monsterExponent}), which is indeed fulfilled. Moreover, Eq.~(\ref{psiNullDB}) can 
be used to calculate $\psi_0$ to third order in $\epsilon$ from our three-loop result 
for $D_B$\cite{janssen_stenull_oerding_99,janssen_stenull_99},
\begin{eqnarray}
\label{backboneExponent}
D_B = 2 + \frac{1}{21} \, \epsilon - \frac{172}{9261} \, \epsilon^2 - 2 \,
 \frac{ 74639 - 22680\, \zeta \left( 3 \right) }{4084101} \, \epsilon^3 + {\sl O}
\left( \epsilon^4  \right) \ .
\end{eqnarray}
We obtain
\begin{eqnarray}
\psi_0 = 1 + \frac{1}{7} \, \epsilon + \frac{313}{12348} \, \epsilon^2 - 
  \frac{ 166823 + 417312\, \zeta \left( 3 \right) }{21781872} \, \epsilon^3 + {\sl O}
\left( \epsilon^4  \right) \ .
\end{eqnarray}
%
%

%
%
\newpage
\thispagestyle{empty}
\epsfxsize=2.3cm
\begin{figure}[h]
\begin{eqnarray*}
\begin{array}{cccccc}
\vspace{5mm}
 & \raisebox{-5mm}{\epsffile{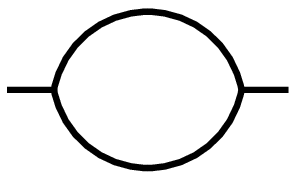}} &=& 
   \raisebox{-5mm}{\epsffile{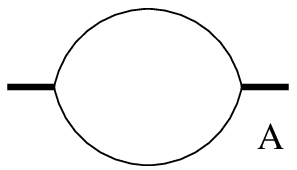}} &-2&  
   \raisebox{-5mm}{\epsffile{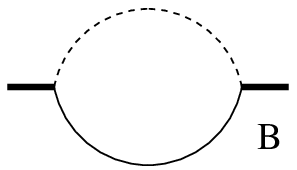}} \\
\vspace{2mm}
 & \raisebox{-5mm}{\epsffile{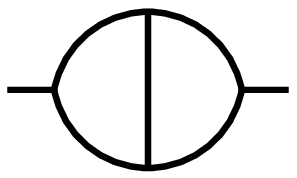}} &=& 
   \raisebox{-5mm}{\epsffile{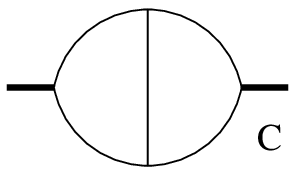}} &-4& 
   \raisebox{-5mm}{\epsffile{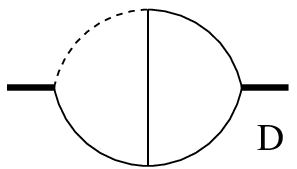}} \\
\vspace{5mm}
-& \raisebox{-5mm}{\epsffile{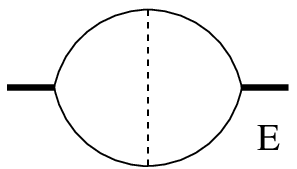}} &+2& 
   \raisebox{-5mm}{\epsffile{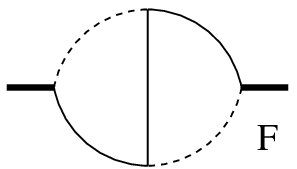}} &+4& 
   \raisebox{-5mm}{\epsffile{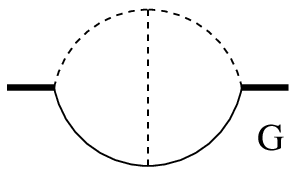}} \\
\vspace{2mm}
 & \raisebox{-5mm}{\epsffile{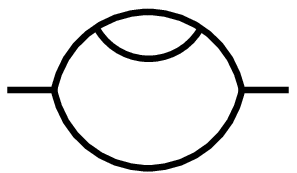}} &=& 
   \raisebox{-5mm}{\epsffile{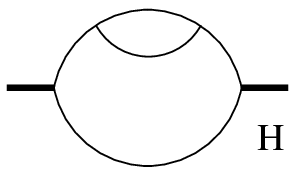}} &-& 
   \raisebox{-5mm}{\epsffile{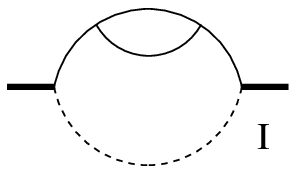}} \\
-2& \raisebox{-5mm}{\epsffile{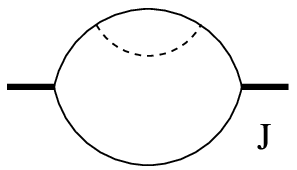}} &+2& 
   \raisebox{-5mm}{\epsffile{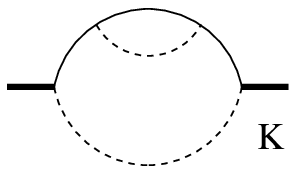}} &+& 
   \raisebox{-5mm}{\epsffile{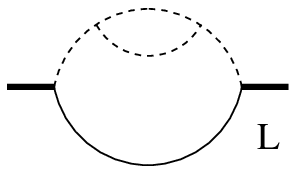}}
\end{array}
\end{eqnarray*}
\caption[]{\label{decomposition}Conducting diagrams to two-loop order. The bold lines symbolize principle propagators, the light lines stand for conducting and the dashed lines for insulating propagators. We point out that the conducting diagrams inherit their combinatorial factor from their bold diagram. For example, the diagrams A and B have to be calculated with the same combinatorial factor $\frac{1}{2}$.}
\end{figure}
\begin{figure}[h]
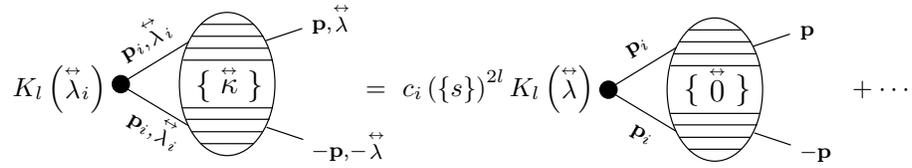

\begin{eqnarray*}
K_l \left( \stackrel{\mbox{{\tiny $\leftrightarrow$}}}{\lambda}_i \right)  
\raisebox{-9mm}{\input{./try.pstex_t}} \hspace{5mm} = \ c_i \left( \left\{ s \right\} 
\right)^{2l} K_l \left( \stackrel{\mbox{{\tiny $\leftrightarrow$}}}{\lambda} \right) 
\raisebox{-10.5mm}{\input{./calcscheme2.pstex_t}} \hspace{5mm} + \cdots
\end{eqnarray*}
\caption[]{\label{calculationScheme}Calculation scheme. The hatched blobs symbolize an arbitrary number of closed conducting loops. The solid dots indicate insertions.}
\end{figure}
\begin{figure}[h]
\begin{eqnarray*}
K_l \left( \stackrel{\mbox{{\tiny $\leftrightarrow$}}}{\lambda}_i \right)  
\raisebox{-9.5mm}{\input{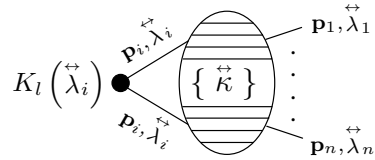}} 
\end{eqnarray*}
\caption[]{\label{reno1}${\mathcal{O}}^{(l)}$ inserted into a diagram with $n$ external legs.}
\end{figure}
\begin{figure}[h]
\begin{eqnarray*}
P_r \left( \stackrel{\mbox{{\tiny $\leftrightarrow$}}}{\lambda} \right) {\rm{\bf 
p}}^{2a}  
\raisebox{-10mm}{\input{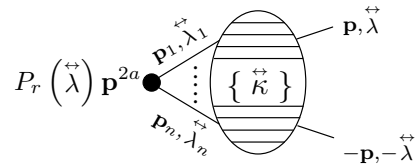}} \hspace{10mm} 
\end{eqnarray*}
\caption[]{\label{reno2}An operator of the type in Eq.~(\ref{generalOperator}) with $n \geq 3$ inserted into
 a two-leg diagram.}
\end{figure}
\begin{figure}[h]
\epsfxsize=7.4cm
\centerline{\epsffile{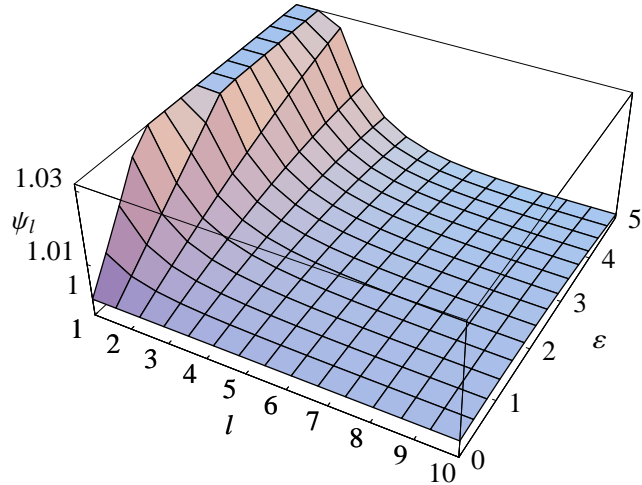}}
\caption[]{\label{3dPlot}Dependence of the noise exponents $\psi_l$ on $\epsilon =6-d$.}
\end{figure}
\begin{figure}[h]
\epsfxsize=10cm
\centerline{\epsffile{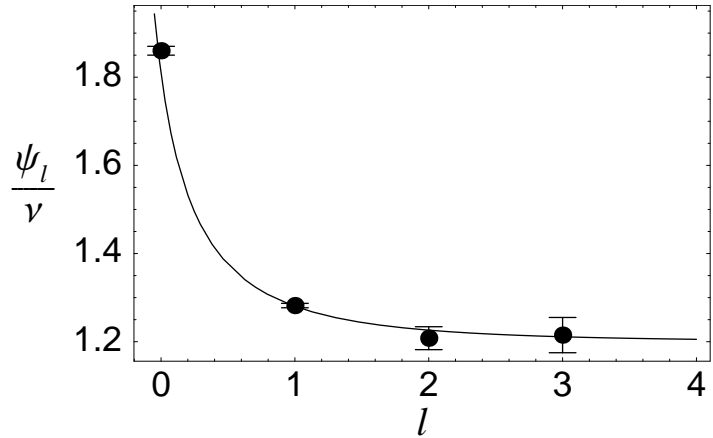}}
\caption[]{\label{psiInDreiD}Dependence $\psi_l /\nu$ on $l$ in three dimensions. The line shows our rationaly approximated $\epsilon$-expansion. The numerical values, symbolized by the dots, are taken from Moukarzel\cite{moukarzel_98} ($l=0$) and Batrouni {\em et al}.\cite{batrouni&co_96} ($l = 1, 2, 3$). Moukarzel determined the backbone dimension $D_B = \psi_0 /\nu$. Batrouni {\em et al}.\ studied the multifractal moments in a fixed voltage ensemble, i.e., for fixed externally applied voltage. The authors state a formula for switching from the multifractal exponents for the fixed voltage ensemble to those for the fixed current ensemble. In this formula a minus sign appears to be missing. Correctly, their exponents $x (n)$ are related to the $\psi_l /\nu$ via $\psi_l /\nu = 2l \, x(2) - x(2l)$.}
\end{figure}

\end{document}